\documentclass[reprint,superscriptaddress,amsmath,amssymb,aps,prb]{revtex4-1}
\usepackage[latin9]{inputenc}
\usepackage{color}
\usepackage{amsmath}
\usepackage{amssymb}
\usepackage{graphicx}

\makeatletter

\usepackage{dcolumn}
\usepackage{bm}\usepackage{graphics}\usepackage{color}\usepackage{epsfig}\usepackage{mathrsfs}
\usepackage{subfigure}
\usepackage{amsfonts}
\usepackage{multirow}
\usepackage{booktabs}\usepackage{float}

\makeatother

\begin{document}
\title{Exact solution of the boundary-dissipated transverse field Ising model:
Structure of Liouvillian spectrum and dynamical duality}
\author{Zhen-Yu Zheng}
\thanks{These authors contributed equally to this work.}
\affiliation{Beijing National Laboratory for Condensed Matter Physics, Institute
of Physics, Chinese Academy of Sciences, Beijing 100190, China}
\author{Xueliang Wang}
\thanks{These authors contributed equally to this work.}
\affiliation{Beijing National Laboratory for Condensed Matter Physics, Institute
of Physics, Chinese Academy of Sciences, Beijing 100190, China}
\affiliation{School of Physical Sciences, University of Chinese Academy of Sciences,
Beijing 100049, China}
\author{Shu Chen}
\thanks{Corresponding author: schen@iphy.ac.cn}
\affiliation{Beijing National Laboratory for Condensed Matter Physics, Institute
of Physics, Chinese Academy of Sciences, Beijing 100190, China}
\affiliation{School of Physical Sciences, University of Chinese Academy of Sciences,
Beijing 100049, China}
\affiliation{Yangtze River Delta Physics Research Center, Liyang, Jiangsu 213300,
China}

\date{\today}
\begin{abstract}
We study the boundary-dissipated transverse field Ising model described by a Lindblad Master equation and exactly solve its Liouvillian spectrum in the whole parameter space.
By mapping the Liouvillian into a Su-Schrieffer-Heeger model with imaginary boundary potentials under
a parity constraint, we solve the rapidity spectrum  analytically and thus construct the Liouvillian spectrum strictly with a parity constraint condition. Our results demonstrate that the Liouvillian spectrum displays four different structures, which
are characterized by different numbers of segments. By analyzing the properties of rapidity spectrum, we can determine the phase boundaries between different spectrum structures analytically and  prove the Liouvillian gap fulfilling a duality relation in the weak and strong dissipation region. Furthermore, we unveil the existence of a dynamical duality, i.e., the long-time relaxation dynamics exhibits almost the same dynamical behavior in the weak and strong dissipation region as long as the duality relation holds true.
\end{abstract}
\maketitle

\section{Introduction}
Advances in quantum engineering of dissipation in laboratory have
attracted a growing interest in the study of open quantum systems
in engineered condensed matter systems \cite{breuer2002,Schindler,weimer},
among which a particularly important class is the boundary-driven
system, where the system is coupled to the environment only at the
boundaries. Within the Markovian approximation, the dynamic evolution
process of a boundary-driven quantum system is governed by the Lindblad
master equation \cite{lindblad1976cmp} with the influence of environment
described by boundary dissipation operators. Understanding dynamical
processes driven by boundary dissipations have attracted intensive
theoretical studies \cite{prosen2009jsm,prosen2008prl,prosen2008njp,prosen2011prl,chuguo2017pra,katsura2019prb,
katsura2020ptep,znidaric2010jsm,znidaric2015pre,zhou2020prb,zhoubz,GuoC2018,chuguo2017pra,yamanaka2021arxiv}.


As a paradigmatic system exhibiting quantum phase transition, the
transverse field Ising model is exactly solvable and has been well
studied in the past decades \cite{sachdev2000,pfeuty1970ap,lieb1961ap,Dutta,katsura}. However, much less is understood
for the corresponding boundary-dissipation-driven model. Recently,
exactly solvable dissipative models have attracted many interests \cite{chuguo2017pra,katsura2019prb,prosen2016prl,yamanaka2021arxiv,GuoC2018,popkov2021prl,nakagawa2021prl,jaksch2020njp}.
Usually, the solvability of these models mainly relies on free-fermion
(boson) techniques or Bethe-ansatz method. One specific class that
has been widely studied is the open quantum systems with quadratic
Lindbladian, which can be solved by third quantization \cite{prosen2008njp,prosen2011prl,yamanaka2021arxiv,GuoC2018,chuguo2017pra,katsura2019prb}.
Although the third quantization method can reduce the problem of solving quadratic
Lindbladian to the diagonalization of a non-Hermitian matrix, analytical solutions are still limited except for some specific cases or for a
special set of parameters \cite{katsura2020ptep,yamanaka2021arxiv,GuoC2018}. The calculation of full Liouvillian spectrum and understanding the spectrum structure in the whole parameter space is still a challenging work.

In this work, we shall present an exact solution to a transverse field
Ising chain with boundary dissipations in the whole parameter space and
construct the Liouvillian spectrum from the rapidity spectrum under the constraint of parity. By vectorizing the density
matrix, solving the Lindblad master equation with boundary dissipation
can be mapped to the solution of the Su-Schrieffer-Heeger (SSH) model
with imaginary boundary potentials \cite{zhu2014pra}, which enables
us to obtain analytical results of the rapidity spectrums.
We stress that the Liouvillian
spectrum can be constructed correctly only when the constraint of parity is properly taken into account. Focusing on the case with equal boundary dissipations, we demonstrate that the Liouvillian spectrum displays four different structures in the whole parameter space. We unveil that the different structures of the Liouvillian
spectrum are determined by number of the complex solutions of equation for solving eigenvalues of the odd-parity rapidity spectrum. The
boundaries between different regions
can be analytically determined via a theoretical analysis in the thermodynamical
limit. 
Furthermore, we prove that the Liouvillian gap fulfills a dual relation in the weak and strong dissipation region and  uncover the existence of a dynamical duality of the relaxation dynamics. Our
work demonstrates novel phenomena of dynamical duality from the perspective of an exact solution and
provides a firm ground for understanding structure of Liouvillian spectrum \cite{popkov2021prl,Zhai}.

The paper is organized as follows. In Sec.II, we rigorously establish the mapping relation between the model and the non-Hermitian matrices associated with parity and emphasize the importance of both even and odd parity. In Sec.III, we solve  the rapidity spectrum  analytically and construct the Liouvillian spectrum via the rapidity spectrum under the parity constraint.
Focusing on the case with equal boundary dissipation strengthes, we demonstrate that the Liouvillian spectrum has four different structures. In Sec. IV, we investigate the Liouvillian gap and the relaxation dynamics  and unveil the existence of a dynamical duality. A summary is given in the last section.

\section{Model and formalism}

We consider the boundary-dissipated open system with the time evolution
of the density matrix $\rho$ described by the Lindblad equation:
\begin{equation}
\frac{d\rho}{dt}=\mathcal{L}[\rho]:=-\mathrm{i}[H,\rho]+\sum_{\mu}(L_{\mu}\rho L_{\mu}^{\dagger}-\frac{1}{2}\{L_{\mu}^{\dagger}L_{\mu},\rho\}).\label{lindblad}
\end{equation}
where we have set $\hbar=1$, and $H$ is the Hamiltonian governing the
unitary part of dynamics of the system described by a transverse field
Ising chain \cite{lieb1961ap,pfeuty1970ap}:
\begin{align}
H=-J\sum_{j=1}^{N-1}\sigma_{j}^{x}\sigma_{j+1}^{x}-h\sum_{j=1}^{N}\sigma_{j}^{z}.\label{H}
\end{align}
Here $N$ is the total number of lattice sites and $\sigma_{j}^{\alpha}(j=1,\cdots,N,\alpha=x,y,z)$
are Pauli matrices at per site. The dissipative processes are described
by the Lindblad operators $L_{\mu}$ with the index $\mu$ denoting
the dissipation channels. Here we consider that the dissipations appear
at left and right edges, i.e.
\begin{align}
L_{L}=\sqrt{\gamma_{L}}\sigma_{1}^{x},~~~L_{R}=\sqrt{\gamma_{R}}\sigma_{N}^{x},\label{LO}
\end{align}
where $\gamma_{L},\gamma_{R}\geq0$ denote the boundary dissipation
strengthes. Here we take $\gamma_{L}=\gamma_{R}=\gamma$ and set $J=1$ as the energy units.

The Liouvillian $\mathcal{L}$ is a superoperator acting in the space
of density matrix operators.
By using the Choi-Jamiolkwski isomorphism\cite{jamiolkowskirmp1972,choilaa1975,tysonjpmg2003,vidal2004prl,kshetrimayum2017nc},
the density matrix is mapped into a vector:
$$\rho=\sum_{mn}\rho_{mn}|m\rangle\langle n|\ \rightarrow\ |\rho\rangle=\sum_{mn}\rho_{mn}|m\rangle\otimes|n\rangle,$$
and thus the Liouvillian can be expressed by the $4^{N}\times4^{N}$
matrix,
\begin{equation}
\begin{aligned}
\mathcal{L}\cong&\mathrm{i}(\textbf{1}\otimes H^{\mathrm{T}}-H\otimes\textbf{1})+\sum_{i}[L_{i}\otimes L_{i}^{*}
-\frac{1}{2}(L_{i}^{\dagger}L_{i}\otimes\textbf{1}\\&+\textbf{1}\otimes L_{i}^{\mathrm{T}}L_{i}^{*})],
\end{aligned}\label{L}
\end{equation}
which gives rise to
\begin{equation}\label{AL2}
\begin{aligned}
\mathcal{L} &=\mathrm{i}(J\sum_{j=1}^{N-1}\sigma_{j}^{x}\sigma_{j+1}^{x}
+h\sum_{j=1}^{N}\sigma_{j}^{z}-J\sum_{j=1}^{N-1}\tau_{j}^{x}\tau_{j+1}^{x} \\&-h\sum_{j=1}^{N}\tau_{j}^{z})+\gamma_{\mathrm{L}}\sigma_{1}^{x}\tau_{1}^{x}
+\gamma_{\mathrm{R}}\sigma_{N}^{x}\tau_{N}^{x}-\left(\gamma_{\mathrm{L}}+\gamma_{\mathrm{R}}\right).
\end{aligned}
\end{equation}
By using the Jordan-Wigner transformation \cite{katsura2020ptep},
\begin{equation}
\begin{aligned}
 &a_{j}=(-1)^{j}\left(\prod_{i=1}^{j-1}\sigma_{i}^{z}\right)\sigma_{j}^{x},\\
 &b_{j}=(-1)^{j}\left(\prod_{i=1}^{j-1}\sigma_{i}^{z}\right)\sigma_{j}^{y},\\
 &\bar{a}_{j}=(-1)^{j}\left(\prod_{i=1}^{N}\sigma_{j}^{i}\right)\left(\prod_{i=1}^{N-j}\tau_{N+1-i}^{z}\right)\tau_{j}^{x},
 \\&\bar{b}_{j}=(-1)^{j-1}\left(\prod_{i=1}^{N}\sigma_{j}^{i}\right)\left(\prod_{i=1}^{N-j}\tau_{N+1-i}^{z}\right)\tau_{j}^{y},
\end{aligned}
\end{equation}
it follows that the Liouvillian $\mathcal{L}$ can be represented
as:
\begin{equation}
  \begin{aligned}
\mathcal{L} &= -h\sum_{j=1}^{N}\left(a_{j}b_{j}+\bar{a}_{j}\bar{b}_{j}\right)
 +J\sum_{j=1}^{N-1}\left(b_{j}a_{j+1}+\bar{b}_{j+1}\bar{a}_{j}\right) \\&+\mathrm{i}\mathcal{P}(\gamma_{\mathrm{L}}\bar{b}_{1}a_{1}+\gamma_{\mathrm{R}}b_{N}\bar{a}_{N})
 -\left(\gamma_{\mathrm{L}}+\gamma_{\mathrm{R}}\right),
   \end{aligned}
\end{equation}
where $a_{j},b_{j},\bar{a}_{j},$ and $\bar{b}_{j}$ are Majorana
fermion operators. Here $\mathcal{P}$ is the parity operator defined
as
\begin{align}
\mathcal{P} & :=\left(\prod_{j=1}^{N}\sigma_{j}^{z}\right)\left(\prod_{j=1}^{N}\tau_{j}^{z}\right).\label{P}
\end{align}
The eigenvalue of the parity operator $\mathcal{P}$ takes a specific
value with $P=\pm1$. It can be checked $[\mathcal{L},\mathcal{P}]=0$.
When $P=-1$, the Liouvillian $\mathcal{L}$ is in the operator space
with odd parity, whereas $P=1$ corresponds to the operator space
with even parity. We notice that the parity operator $\mathcal{P}$
gives a strong constraint on the mapping between the spin Hilbert
space and the fermionic Hilbert space. Because the total degrees of
freedom of the Liouvillian with odd parity $\mathcal{L}^{P}|_{P=-1}$
and even parity $\mathcal{L}^{P}|_{P=1}$ is twice as the degrees
of freedom of the Liouvillian $\mathcal{L}$, this gives rise to redundant
degrees of freedom. To eliminate the redundant degrees of freedom,
we apply the projection and the mapping relation as follows,
\begin{equation}
\mathcal{L}=\mathrm{P}^{+}\mathcal{L}^{P}|_{P=1}\mathrm{P}^{+}
+\mathrm{P}^{-}\mathcal{L}^{P}|_{P=-1}\mathrm{P}^{-},\label{mapping}
\end{equation}
where the projectors are defined as
\begin{equation}
\mathrm{P}^{\pm}=\frac{1}{2}(1\pm\mathcal{P}).\label{projectors}
\end{equation}

For convenience, we use the creation operator and the annihilation
operator of fermions, defined as
\begin{align}
a_{j}=c_{2j-1}+c_{2j-1}^{\dagger}, & \quad\bar{b}_{j}=\frac{1}{\mathrm{i}}\left(c_{2j-1}-c_{2j-1}^{\dagger}\right)\\
\bar{a}_{j}=c_{2j}+c_{2j}^{\dagger}, & \quad b_{j}=\frac{1}{\mathrm{i}}\left(c_{2j}-c_{2j}^{\dagger}\right),
\end{align}
to rewrite the Liouvillian as
\begin{align}
 \mathcal{L}^{P}
& =  2\mathrm{i}\left[h\sum_{j=1}^{N}(c_{2j-1}^{\dagger}c_{2j}+\text{h.c.})+J\sum_{j=1}^{N-1}(c_{2j}^{\dagger}c_{2j+1}
+\text{h.c.})\right] \nonumber \\
 & -2\gamma_{\mathrm{L}}P(c_{1}^{\dagger}c_{1}-\frac{1}{2})-2\gamma_{\mathrm{R}}(c_{2N}^{\dagger}c_{2N}-\frac{1}{2})-(\gamma_{\mathrm{L}}+\gamma_{\mathrm{R}}) \nonumber \\
& = 2\mathrm{i}\mathbf{c}^{\dagger}\mathrm{T}^{P}\mathbf{c}+\gamma_{\mathrm{L}}(P-1),
\end{align}
where the spinors are denoted as $\mathbf{c}^{\dagger}=\left(c_{1}^{\dagger},\ldots,c_{2N}^{\dagger}\right),\mathbf{c}=\left(c_{1},\ldots,c_{2N}\right)^{\mathrm{T}}$
and $\mathrm{T}^{P}$ is represented in terms of a $2N\times2N$ non-hermitian
matrix as follow,
\begin{align}
\mathrm{T}^{P}=\left(\begin{array}{cccccc}
P\mathrm{i}\gamma_{\mathrm{L}} & h & \cdots & \cdots &  & 0\\
h & 0 & J\\
 & J & 0 & \ddots\\
 &  & \ddots & \ddots & \ddots\\
 &  &  & \ddots & 0 & h\\
0 &  & \cdots & \cdots & h & \mathrm{i}\gamma_{\mathrm{R}}
\end{array}\right).\label{ATp}
\end{align}
In terms of fermion operator, the parity operator is represented as
$
\mathcal{P}=(-1)^{\hat{N}_f}
$
with $\hat{N}_f=\sum_{j=1}^{2N}c_{j}^{\dagger}c_{j}$ being the total complex
fermion number operator.
The parity of the Liouvillian
$\mathcal{L}$ corresponds to the total number of complex fermions being even or odd, with $P=1$ and $-1$ corresponding to the even and odd parity, respectively.
For the case with $\gamma_L=\gamma_R=\gamma$, the Liouvillian with a specific parity is written as
\begin{equation}
 \mathcal{L}^{P} = 2\mathrm{i}\mathbf{c}^{\dagger}\mathrm{T}^{P}\mathbf{c}+\gamma(P-1),
\end{equation}
where $\mathrm{T}^{P}$ is given by
\begin{align}
\mathrm{T}^{P}=\left(\begin{array}{cccccc}
P\mathrm{i}\gamma & h & \cdots & \cdots &  & 0\\
h & 0 & 1\\
 & 1 & 0 & \ddots\\
 &  & \ddots & \ddots & \ddots\\
 &  &  & \ddots & 0 & h\\
0 &  & \cdots & \cdots & h & \mathrm{i}\gamma
\end{array}\right).\label{Tp}
\end{align}
Here we have set $J=1$. It is clear that $\mathrm{T}^{P}$ describes a non-Hermitian SSH model \cite{su1979prl} with imaginary
boundary potentials \cite{zhu2014pra,klett2017pra,katsura2020ptep}.

By using
the eigen-decomposition $\mathrm{T}^{P}=\sum_{j=1}^{2N}E_{j,P}|\Psi_{j,P}\rangle\langle\Phi_{j,P}|$,
we can get the diagonal form of the Liouvillian:
\begin{equation}
\mathcal{L}^{P}=2\mathrm{i}\sum_{j=1}^{2N}E_{j,P}\overline{d}_{j,P}d_{j,P}+\gamma (P-1),\label{LTD}
\end{equation}
where $d_{j,P}=\sum_{i=1}^{2N}\zeta_{j,P,i}c_{i}$ and $\overline{d}_{j,P}=\sum_{i=1}^{2N}\xi_{j,P,i}c_{i}^{\dagger}$.
The parameters $\xi_{j,P,i}$ and $\zeta_{j,P,i}$ are the $i$-th element of $|\Psi_{j,P}\rangle$
and $\langle\Phi_{j,P}|,$ respectively. Here we take the Bogoliubov
modes as $\left(d_{j,P},\overline{d}_{j,P}\right)$ instead of $\left(d_{j,P},d_{j,P}^{\dagger}\right)$,
which satisfy the canonical anti-commutation relations\cite{song2013pra,song2014pra}
\begin{align}
 & \left\{ d_{j,P},\overline{d}_{j^{\prime},P}\right\} =\delta_{jj^{\prime}}, \nonumber \\
 & \left\{ d_{j,P},d_{j^{\prime},P}\right\} =\left\{ \overline{d}_{j,P},\overline{d}_{j^{\prime},P}\right\} =0. \nonumber
\end{align}
According to Eq.(\ref{mapping}),
the eigenstates of the Liouvillian $\mathcal{L}$ comes from two parts
which contain all occupied states of  even complex fermions  from $\mathcal{L}^{P}|_{P=1}$
and of   odd complex fermions from $\mathcal{L}^{P}|_{P=-1}$.
Thus, we can get the full eigenstates and the spectrum of Liouvillian $\mathcal{L}$ by the reorganizations of the rapidity spectrum of $\mathcal{L}^{P}|_{P=1}$ and $\mathcal{L}^{P}|_{P=-1}$.
\begin{figure*}[t]
\center
\includegraphics[width=17cm]{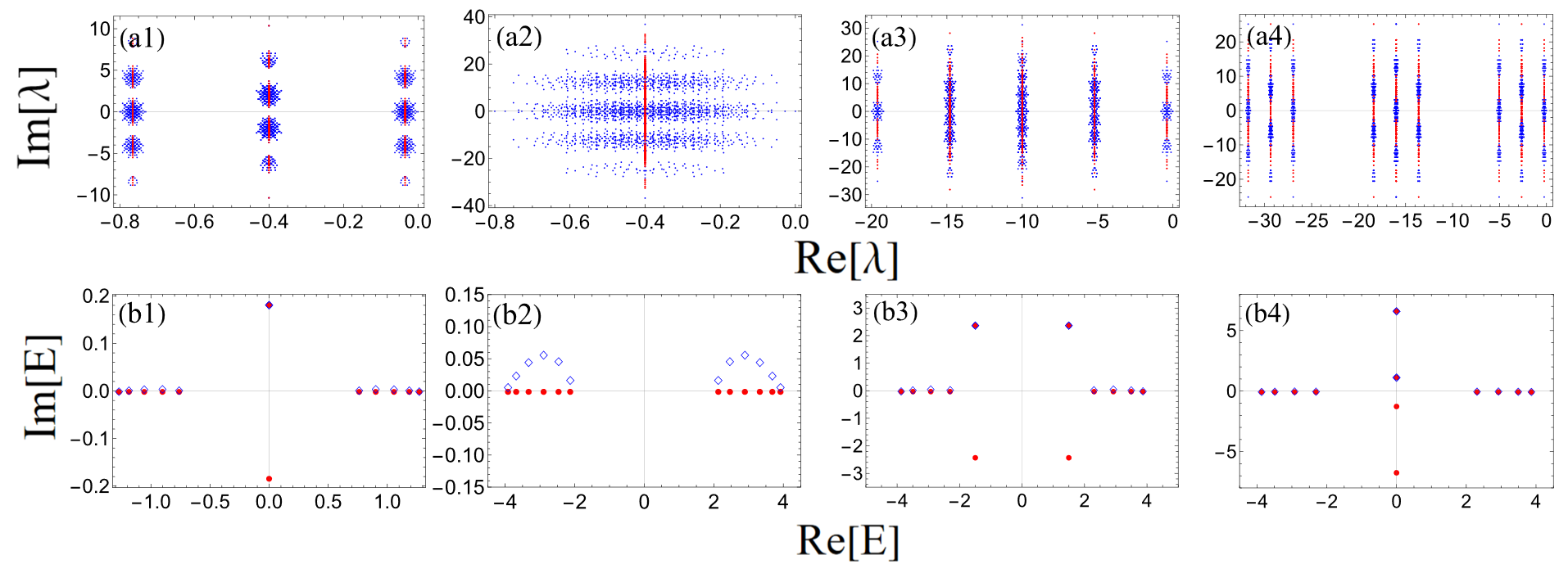} \caption{The Liouvillian spectrum and the rapidity spectrum with $N=6$, and
(a1), (b1) $h=0.3,\gamma=0.2$, and (a2), (b2) $h=3,\gamma=0.2$
, and (a3), (b3) $h=3,\gamma=5$ and (a4), (b4) $h=3,\gamma=8$.
The red points represent the Liouvillian spectrum of odd channel with $P=-1$,
while the blue ones represent that of even channel with $P=1$. The
Liouvillian spectrum in (a1), (a2), (a3), and (a4) can be constructed
by the rapidity spectrum in (b1), (b2), (b3), and (b4), respectively.
The eigenvalues of Liouvillian spectrum in (a1), (a2), (a3), and (a4)
satisfy $\Re[\lambda]\protect\leq0$
and the data of Liouvillian spectrum are consistent with ones by exact
diagonalization. The red points in (b1), (b2), (b3), and (b4) represent the rapidity spectrum from odd channel and the blue empty prisms represent the rapidity spectrum from even channel.}
\label{FLS1}
\end{figure*}

\section{Structure of Liouvillian spectrum}

The full spectrum of Liouvillian $\mathcal{L}$ can be obtained by
reorganizing the rapidity spectrum of $\mathcal{L}^{P}|_{P=1}$ and
$\mathcal{L}^{P}|_{P=-1}$, which can be analytically derived by solving
the eigenvalues of $\mathrm{T}^{P}$, i.e.,
\begin{equation}
\mathrm{T}^{P} \Psi_{P} =E_{P} \Psi_{P} , \label{Teq}
\end{equation}
where $\Psi_{P} $ denotes the eigenvector corresponding to the eigenvalue $E_{P}$ (the rapidity spectrum).
We can exactly solve the eigenvalues by applying the analytical method in Ref.\cite{GuoCX}. In terms of the parameter $\theta$, the eigenvalue can be represented
as
\begin{equation}
E_{P}=\pm\sqrt{1+h^{2}+2h\cos\theta}.\label{E}
\end{equation}
The value of $\theta$ is determined by the boundary equations \cite{GuoCX} ,
which leads to the following equation (the details are shown in Appendix A):
\begin{align}
p_{1}\sin[N\theta]-p_{2}\sin[(N+1)\theta]+p_{3}\sin[(N-1)\theta]=0,\label{theta}
\end{align}
where $p_{1}=\mathrm{i}(P+1)\gamma E_{P}-(1-P\gamma^2)$,
$p_{2}=h$, and $p_{3}=hP\gamma^2$.

By solving Eq.(\ref{theta}), we can obtain the value of $\theta_{j}$
and thus the rapidity spectrum.
Explicitly, we rewrite Eq.(\ref{theta})
in even channel ($P=1$) and odd channel ($P=-1$) as
\begin{equation}
\begin{array}{c}
[2\mathrm{i}\gamma E_{e}-(1-\gamma^{2})]\sin[N\theta]-h\sin[(N+1)\theta]\\
+h\gamma^{2}\sin[(N-1)\theta]=0
\end{array}\label{eq:thetaN1}
\end{equation}
and
\begin{equation}
(1+\gamma^{2})\sin[N\theta]+h\sin[(N+1)\theta]+ h\gamma^{2}\sin[(N-1)\theta]=0, \label{eq:thetaN2}
\end{equation}
respectively. For convenience, we denote the solutions of Eq.(\ref{eq:thetaN1})
as $\theta_{e}$ and of Eq.(\ref{eq:thetaN2}) as $\theta_{o}$, respectively. Substituting them into the formula
of eigenvalue in Eq.(\ref{E}), we can get the rapidity spectrum, denoted as $E_{j,e}$
and $E_{j,o}$ which are eigenvalues
of $\mathrm{T}^{e}$ and $\mathrm{T}^{o}$, respectively, with $j=1,\cdots,2N$. By considering
the constraint of the parity operator and Eq.(\ref{mapping}), the full spectrum of the Liouvillian $\mathcal{L}$
can be exactly expressed as
\begin{align}
\lambda=\left\{ \begin{array}{ll}
2\mathrm{i}\sum_{j=1}^{2N}v_{j,e}E_{j,e}~~ & (v_{j,e}=0,1),\\
2\mathrm{i}\sum_{j=1}^{2N}v_{j,o}E_{j,o}-2\gamma ~~ & (v_{j,o}=0,1),
\end{array}\right.\label{ZHE}
\end{align}
where $\sum_{j=1}^{2N}v_{j,e}$ is even and $\sum_{j=1}^{2N}v_{j,o}$
is odd. The constraint on the total complex fermion number
removes the redundant degrees of freedom.

\begin{figure}[t]
\centering \includegraphics[width=7.5cm]{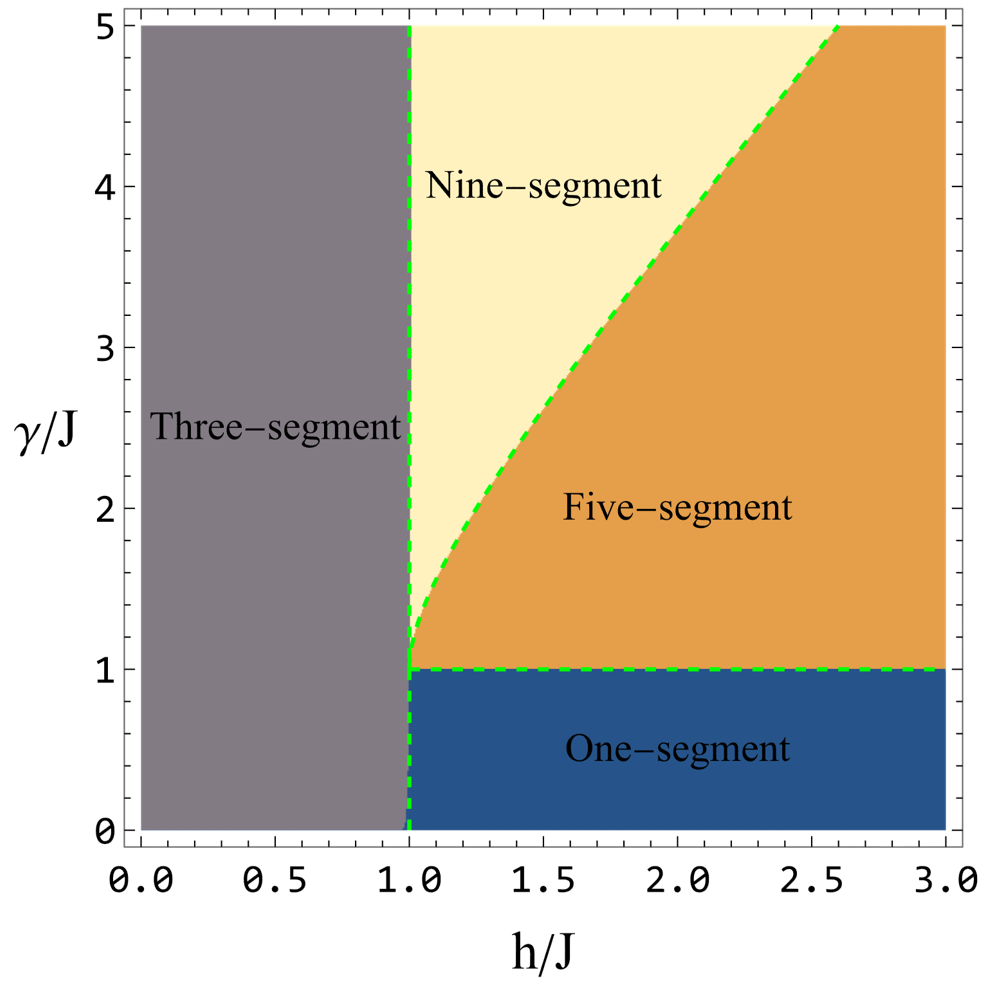} \caption{The schematic phase diagram for the stucture of the Liouvillian spectrum. The green dashed
lines denote the boundaries between different regions with different number of eigenvalue segments.}
\label{syt}
\end{figure}

In Fig.\ref{FLS1} we demonstrate the Liouvillian spectrum and the
corresponding rapidity spectrum for four typical cases. The rapidity
spectrum is obtained by numerically solving Eq.(\ref{eq:thetaN1})
and Eq.(\ref{eq:thetaN2}) and thus the Liouvillian spectrum is obtained
from Eq.(\ref{ZHE}).  The Liouvillian spectrum displays
different structures in the four parameter regions, as schematically displayed in Fig.\ref{syt}.  We have checked our Liouvillian spectra by comparing with the numerical results via the diagonalization of Liouvillian and find that they agree exactly.

\begin{figure}[t]
\centering \flushleft \includegraphics[width=8.5cm]{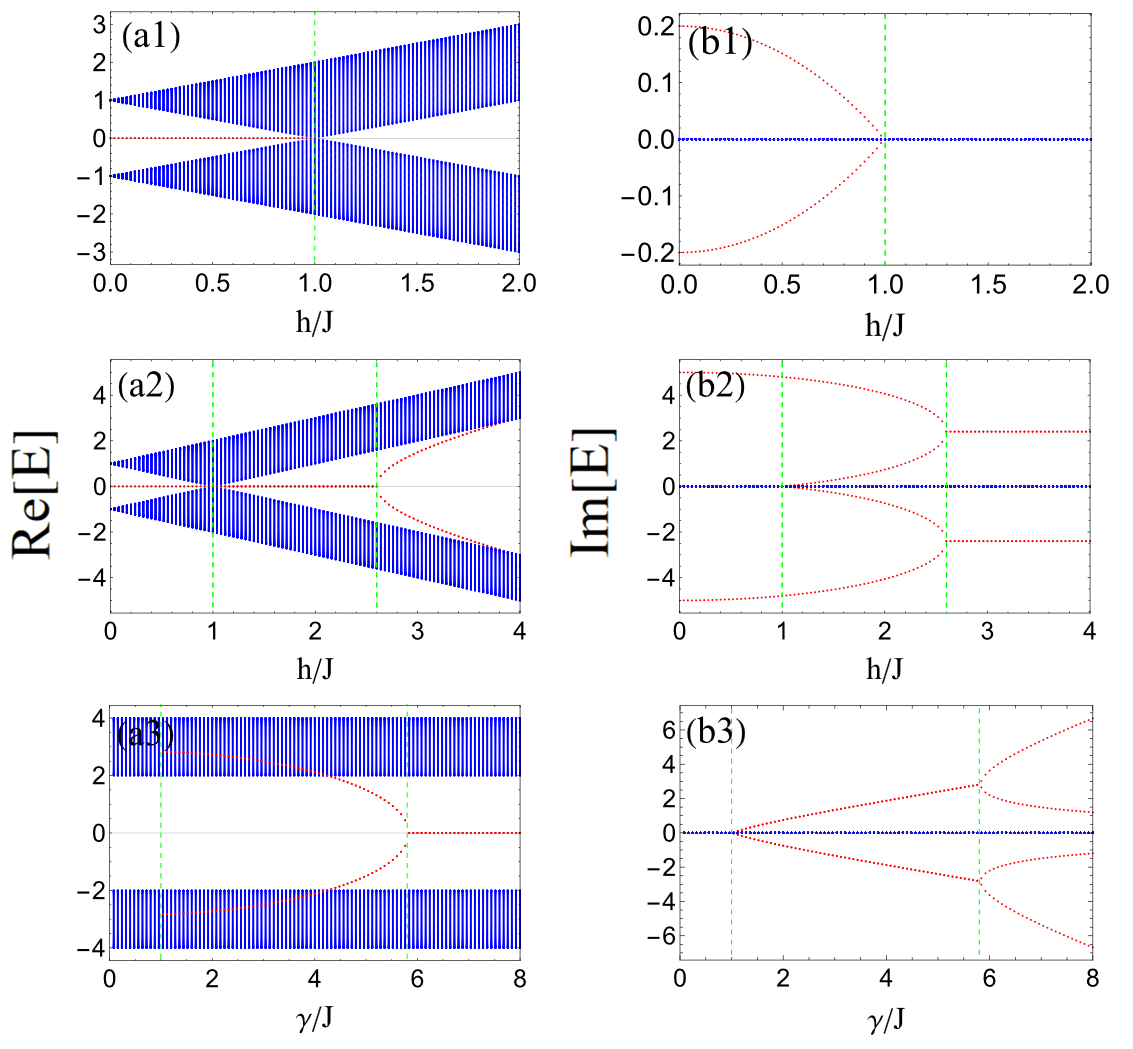} \caption{The rapidity spectrum from the odd channel.
The red lines denotes the non-real roots and green dash lines indicate the boundaries of regions with different complex solutions. We set the parameters with (a1),
(b1) $N=100,\gamma=0.2$, (a2), (b2) $N=100,\gamma=5$ and
(a3), (b3) $N=100,h=3$.}
\label{rps}
\end{figure}

We observe that the Liouvillian spectrums from the odd channel present
distinct stripes and from the even channel
are scattered near these stripes, as shown in Figs.\ref{FLS1}(a1)-(a4). For convenience, we call one stripe and points surrounding this stripe as one segment. The distance between each segment is determined by the imaginary part of rapidity spectrum of the odd channel,
and the width of the segments is determined by the imaginary part of rapidity spectrum of the even channel close to the real axis.
The number of segments is closely related to number of complex solutions of the rapidity spectrum of the odd channel, which correspond to the boundary bound states of $T^{o}$ \cite{zhu2014pra}.
Since $T^{o}$ fulfills $\mathscr{P}\mathcal{T}$ (parity and time-reversal) symmetry, solutions of Eq.(\ref{eq:thetaN2}) are either real or occur in complex conjugated pairs. In the $\mathscr{P}\mathcal{T}$-symmetry region of $h>1$ and $\gamma<1$, all $N$ solutions of Eq.(\ref{eq:thetaN2}) are real. The corresponding rapidity spectrum has no pure imaginary eigenvalues, and the Liouvillian spectrum displays a structure composed of one segment. In the region of $h<1$,  the odd rapidity spectrum has one pair of purely imaginary eigenvalues (see  Fig.\ref{FLS1}(b1) and Figs.\ref{rps}(a1) and (b1)), and the
Liouvillian spectrum is composed of three segments. For $h>\frac{1+\gamma^{2}}{2\gamma}$ and $\gamma>1$,
the odd rapidity spectrum has two conjugated pairs of complex
eigenvalues which are symmetrical about the imaginary axis (see Fig.\ref{FLS1}(b3) and Figs.\ref{rps}(a3) and (b3)), and the Liouvillian spectrum displays a structure of five segments.
For $h=3$ and $\gamma=8$, the odd rapidity spectrum has two conjugated pairs
of purely imaginary eigenvalues, inducing that the Liouvillian spectrum
presents a structure of nine segments.

For the even channel, $T^{e}$ fulfills the reflection symmetry and $K$ symmetry. The corresponding solutions of Eq.(\ref{eq:thetaN1}) are complex and distribute symmetrically about the imaginary axis (see Appendix A for details). As shown in  Fig.\ref{FLS1}(b1)-(b4), the rapidity spectrum from the even channel has a one-to-one correspondence to the spectrum from the odd channel. For the eigenvalues close to the real axis, their imaginary part determines the width of the segments in the Liouvillian spectrum.
There also exist complex eigenvalues farther away from the real axis, which are degenerate and almost overlap with one (ones) of the corresponding complex conjugated pairs (in the upper half-plane) from the odd channel, as shown in Figs.\ref{FLS1}(b1), (b3) and (b4). Similarly, the number of segments is determined by the number of
this kind of complex solutions, which correspond to the boundary bound states of $T^{e}$.

\begin{figure}[t]
\center \includegraphics[width=8.5cm]{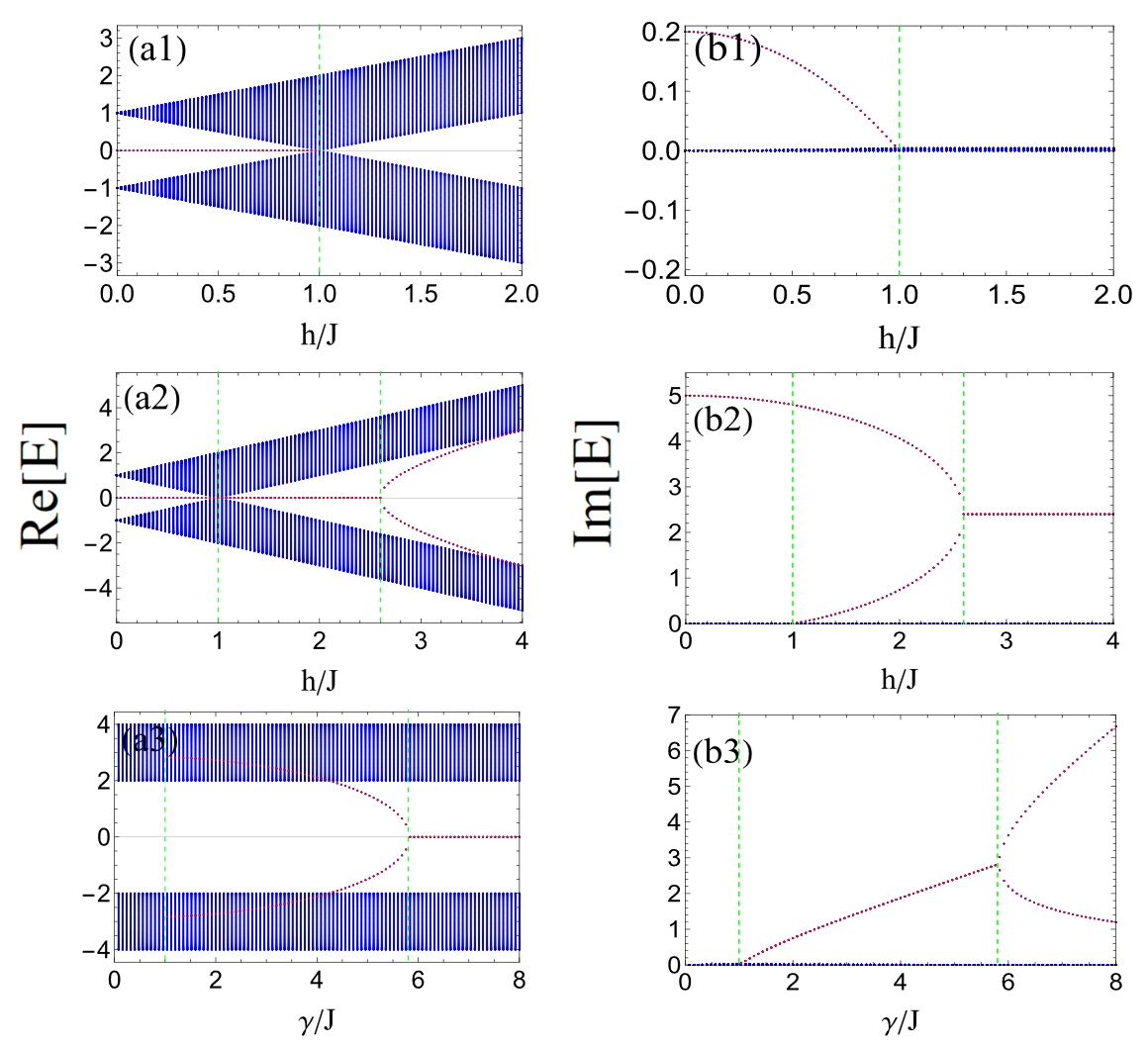} \caption{The rapidity spectrum from the even channel.
The red lines denote the non-real roots  corresponding to  the boundary bound states and green dash lines indicate the boundaries of regions with different complex solutions. We set the parameters with (a1),
(b1) $N=100,J=1,\gamma=0.2$, (a2), (b2) $N=100,J=1,\gamma=5$ and
(a3), (b3) $N=100,J=1,h=3$.}
\label{appendix3}
\end{figure}

As shown in Fig.\ref{syt},  different structures of Liouvillian
spectrum are characterized by different numbers of segments in four regions.
Boundaries of phases with different spectrum
structures can be determined by boundaries of parameter regions with different numbers of complex rapidity eigenvalues.
To see it clearly, in Fig.\ref{rps}, we show
the change of the odd-parity rapidity spectrum with the parameter $h$ (by fixing $\gamma=0.2$ and $5$, respectively) and $\gamma$ (by fixing $h=3$), for the system $N=100$.
In the thermodynamic limit, we can  analytically determine the boundaries of regions with different spectrum structures (the details are shown in Appendix B).
The odd-parity rapidity spectrum has one pair of pure imaginary eigenvalues in the region with $h<1$, two pairs of pure imaginary eigenvalues in the region with $h>1$, $\gamma>1$ and $h<\frac{1+\gamma^{2}}{2\gamma}$, no pure imaginary eigenvalue in the region with $h>1$ and $\gamma<1$, and two pairs of complex eigenvalues in the region with $h>\frac{1+\gamma^{2}}{2\gamma}$ and $\gamma>1$.

As a comparison, we also demonstrate the even-channel rapidity spectrum.
To see clearly how the rapidity spectrum changes with parameters, in Fig.\ref{appendix3}, we show
the change of the even-parity rapidity spectrum with the parameter $h$ (by fixing $\gamma=0.2$ and $5$, respectively) and $\gamma$ (by fixing $h=3$), for the system $N=100$. Compared with the ones from the odd-parity rapidity spectrum in Fig.3, the real part of the boundary bound states is the same as the ones of the odd-parity rapidity spectrum and the imaginary part of them is only half of the ones of the odd-parity rapidity spectrum. Here the boundary bound states are doubly degenerate. It is shown that the boundaries of different structures of Liouvillian spectrum can be also obtained from the even-channel rapidity spectrum.

\section{Liouvillian gap and relaxation dynamics}
Next we discuss the Liouvillian
gap $\Delta_{g}$,
which is given by the eigenvalue with the
largest nonzero real part, i.e.,
$
\Delta_{g}:=-\max~\Re[\lambda]|_{\Re[\lambda]\neq0}
$  \cite{znidaric2015pre,CaiZ}.
Explicitly, the Liouvillian gap can be represented as
\begin{equation}
\Delta_{g}=-\Re[2\mathrm{i}(E_{j_{1},e}+E_{j_{2},e})],\label{dg}
\end{equation}
where $E_{j_{1},e}$ and $E_{j_{2},e}$ are two eigenvalues with minimum imaginary part in the rapidity spectrum from the even
channel. As shown in Fig.1(b1)-(b4), the eigenvalues always distribute symmetrically about the imaginary axis, i.e, $E_{j_{1},e}=-E^*_{j_{2},e}$ due to the $K$ symmetry.
We note that the sum of $E_{j_{1},e}$ and $E_{j_{2},e}$ in Eq.(\ref{dg}) is due to the constraint of parity. If the constraint is not properly accounted, the Liouvillian gap is underestimated and takes only half of the value of $\Delta_{g}$.

\begin{figure}[h]
\centering\includegraphics[width=8cm]{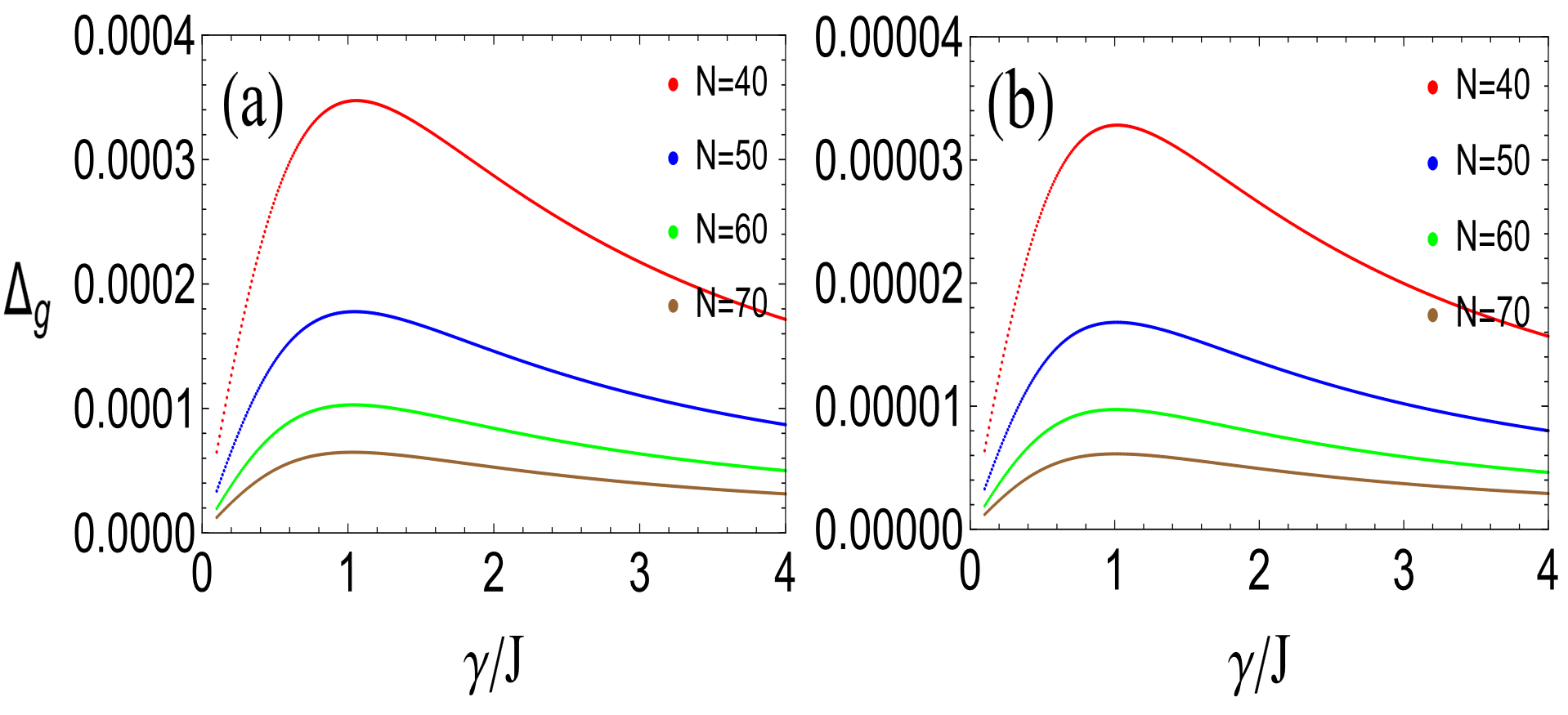} \caption{The Liouvillian gap with (a) $J=1$, $h=0.3$ and (b) $J=1$, $h=3$.}
\label{Lgap}
\end{figure}

In the weak and strong dissipation limit, we can derive an analytical expression for the Liouvillian gap by applying a
perturbative expansion in terms of the small parameter $\gamma$ or $1/\gamma$, which leads to $\Delta_{g} \propto\gamma N^{-3}$ for $\gamma\ll 1$ and
 $\Delta_{g}\propto\frac{N^{-3}}{\gamma}$ for $\gamma \gg 1$ (the details are shown in Appendix C).
In the thermodynamic limit, we can prove the Liouvillian gap fulfills a dual relation
\begin{equation}
  \Delta_{g}(\gamma,h)= \Delta_{g}(\frac{1}{\gamma},h),
\end{equation}
which holds true for arbitrary $\gamma$ and is irrelevant to $h$ (see Appendix D for the proof).
From the dual relation, we can conclude that the Liouvillian gap takes its maximum at $\gamma=1$ in the whole parameter region of $h$.
As shown in Fig.\ref{Lgap}, our numerical results also confirms that the Liouvillian gap
increases with the increase of $\gamma$ in the regime of weak dissipation whereas decreases with the increase of $\gamma$ in the region of strong dissipation. From the numerical results for system with different sizes, it can be inferred that the inflection
point is at $\gamma=1$ when $N$ tends to infinity, consistent with the duality relation obtained in the thermodynamical limit.

The duality relation of Liouvillian gap also suggests that the relaxation times in the weak ($\gamma \ll 1)$ and strong dissipation regions ($\gamma'=1/\gamma \gg 1$) should be the same.
Furthermore, we find that the most of rapidity spectrum satisfies the duality relation $E(\gamma)=E(\frac{1}{\gamma})$ in the thermodynamic limit (see Appendix D), except of those corresponding to the bound states, which contribute to the distance between segments of Liouvillian spectrum. The existence of such a duality relation
means that the rightmost segment of Liouvillian spectrum (the one close to the steady state) in the weak ($\gamma \ll 1)$ and strong dissipation regions ($\gamma'=1/\gamma \gg 1$) are almost the same. So we can predict that the system in the weak and strong regions should display almost the same relaxation dynamics when the evolution time enters the region dominated by the rightmost segment, i.e., the existence of a \emph{dynamical duality} in the weak and strong dissipation regions.

To get an intuitive understanding, next we investigate the dynamical behaviour by calculating the time-dependent average magnetization denoted as
\begin{align}
\langle m^{z}(t)\rangle=\langle\frac{1}{N}\sum_{i=1}^{N}\sigma_{i}^{z}(t)\rangle.\label{mz}
\end{align}
The initial state is assumed as $|\rho_{0}\rangle$ and the vectorizing
form of the initial state is $|\rho_{0}\rangle=\text{vec}(\rho_{0})$.
Then the time-dependent state can be denoted as $\left|\rho(t)\right\rangle =e^{\mathcal{L}t}\left|\rho_{0}\right\rangle $,
where the eigen-expansion form of $\mathcal{L}$ is performed as $\mathcal{L}=\sum_{i}\lambda_{i}\left|\psi_{i}^{r}\right\rangle \left\langle \psi_{i}^{l}\right|$.
Then, the average magnetization is rewritten as
\begin{equation}
\begin{aligned}
\left\langle m^{z}(t)\right\rangle & =\text{Tr}[m^{z}(t)\rho(t)]=\left\langle m^{z}(t)\right|\left.\rho(t)\right\rangle
\\&=\sum_{i}e^{\lambda_{i}t}\left\langle m^{z}(t)\right|\left.\psi_{i}^{r}\right\rangle \left\langle \psi_{i}^{l}\right|\left.\rho_{0}\right\rangle,
\end{aligned}
\end{equation}
where $\left\langle m^{z}(t)\right|\equiv[\text{vec}(m^{z}(t)^{\dagger})]^{\dagger}$.
Here,  we choose the state with all spin up as the initial state.

\begin{figure}[t]
\centering\includegraphics[width=8.5cm]{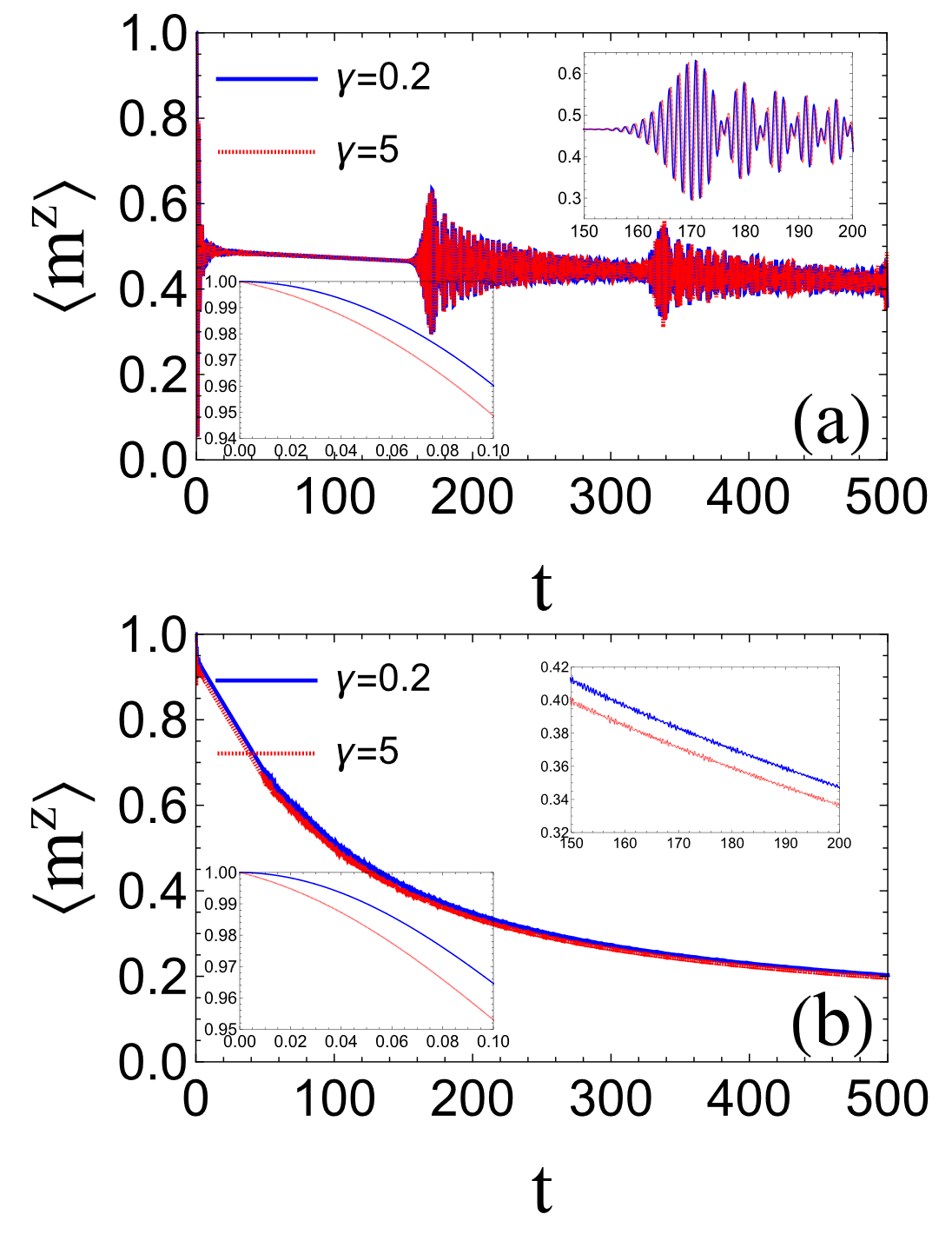} \caption{The dynamical evolution of the average magnetization with $N=100$, (a) $h=0.3$ and (b) $h=3$.}
\label{Dynamic}
\end{figure}

Because the Liouvillian of our model is quadratic form, alternatively we can calculate the average magnetization by using Lyapunov equation method \cite{zhang2022arxiv}  (see Appendix E), which enable us to calculate the dynamics of systems with large sizes.
We calculate the time-dependent average magnetization
in both the weak and strong dissipation regions. By choosing the state with all spin up as the initial state, we display the time evolution of the average magnetization for various parameters in Fig.\ref{Dynamic}.
It is shown that the relaxation dynamics for systems with $\gamma=0.2$ and $5$ are almost identical except in very short time.
As the short-time dynamics is mainly determined by the leftmost segment of Liouvillian spectrum, whose center position is determined by the boundary bound state, at the beginning time $\langle m^{z}(t)\rangle$ decays more slowly for the case with $\gamma=0.2$ than that with $\gamma=5$ as shown in the left insets of Fig.\ref{Dynamic}.

\section{Summary}

We have exactly solved the transverse field Ising model with boundary
dissipations described by the Lindblad master equation. Under a parity constraint, the Liouvillian spectrum is constructed
strictly via the rapidity spectrum from both odd and even channels. We find
 the Liouvillian spectrum displaying four different
structures in the whole parameter space and determine the phase boundaries of different structures analytically in the thermodynamical limit. Our analytical results also unveil that the Liouvillian gap fulfills a duality relation in the weak and strong dissipation region and the relaxation dynamics also exhibits a dynamical duality.

Our analytical results unveil that the number of stripes is closely related to the number of complex boundary states in the odd-parity rapidity spectrum. Therefore, we expect that the stripe structures are stable against perturbations. To verify this, we introduce random on-site disorder perturbation in the transverse field of Hamiltonian (2) and numerically calculate the corresponding Liouvillian spectrum. Our numerical results show that the stripe structures are stable even the perturbed disorder strength is about ten percent of the strength of transverse field. It is also interesting to explore to what extent the dynamical duality exists in other open systems in future works.

\begin{acknowledgments}
We thank C.-X. Guo and C. G. Liang for useful discussions. The work is supported by the NSFC under Grants No.12174436
and No.T2121001 and the Strategic Priority Research Program of Chinese Academy of Sciences under Grant No. XDB33000000.
\end{acknowledgments}

\appendix

\section{Analytical solution of the rapidity spectrum}

The full spectrum of Liouvillian $\mathcal{L}$ can be obtained by
reorganizing the rapidity spectrum of $\mathcal{L}^{P}|_{P=1}$ and
$\mathcal{L}^{P}|_{P=-1}$, which can be analytically derived by solving
the eigenvalues of the matrix $\mathrm{T}^{P}$. We consider the general
case with $J\neq0$ and $h\neq0$ and solve the eigenvalue equation
\begin{equation}
\mathrm{T}^{P} \Psi_{P} =E_{P} \Psi_{P} \label{BTeq}
\end{equation}
by following the analytical method in Ref.\cite{GuoCX}, where we
denote $\Psi_{P}=\left(\psi_{1,A},\psi_{1,B},\psi_{2,A},\cdots,\psi_{N,B}\right)^{T}$.

By substituting Eq.(\ref{ATp}) into Eq.(\ref{BTeq}), we get of a series
of bulk equations
\begin{equation}
\begin{array}{c}
J\psi_{(n-1),B}+h\psi_{n,B}-E_{P}\psi_{n,A}=0,\\
h\psi_{(n-1),A}+J\psi_{n,A}-E_{P}\psi_{(n-1),B}=0,
\end{array}\label{eq:Ebulk}
\end{equation}
with $n=2,\cdots,N$, and the boundary equations given by
\begin{equation}
\begin{array}{c}
P\mathrm{i}\gamma_{L}\psi_{1,A}+h\psi_{1,B}-E_{P}\psi_{1,A}=0,\\
h\psi_{N,A}+\mathrm{i}\gamma_{R}\psi_{N,B}-E_{P}\psi_{N,B}=0.
\end{array}\label{eq:Eboundry}
\end{equation}
Due to the spatial translational property of bulk equations, we set
the ansatz of wave function $|\Psi_{P}\rangle$ as follows,
\begin{equation}
\Psi_{P} =\left(z\phi_{A},z\phi_{B},z^{2}\phi_{A},z^{2}\phi_{B},\cdots,z^{N}\phi_{A},z^{N}\phi_{B}\right)^{T}.\label{waveF}
\end{equation}
By substituting it into the bulk equations Eq.(\ref{eq:Ebulk}), we
obtain
\begin{equation}
Jhz^{2}+(J^{2}+h^{2}-E_{P}^{2})z+Jh=0.\label{Ebulksp}
\end{equation}
For a given $E_{j,P}$, there are two solutions $z_{i}\left(z_{1},z_{2}\right)$.
According to Vieta's theorem, we can get two constraint equations
of them from Eq.(\ref{Ebulksp})
\begin{eqnarray}
  z_{1}+z_{2}=\frac{E_{P}^{2}-J^{2}-h^{2}}{Jh}, z_{1}z_{2}=1.\label{zeq}
\end{eqnarray}
The constraint condition of Eq. (\ref{zeq}) suggests that the solutions
can be represented as
\begin{equation}
z_{1}=e^{i\theta},\quad z_{2}=e^{-i\theta}.\label{ze}
\end{equation}
In terms of the parameter $\theta$, the eigenvalue can be represented
as
\begin{equation}
E_{P}=\pm\sqrt{J^{2}+h^{2}+2Jh\cos\theta}.\label{BE}
\end{equation}
The value of $\theta$ shall be determined by the boundary equations.

Since the superposition of two linearly independent solutions is also
the solution of bulk equations, the general wave function takes the
form of
\begin{align}
\psi_{n,A}=g_{1}z_{1}^{n}\phi_{A}^{(1)}+g_{2}z_{2}^{n}\phi_{A}^{(2)},\label{gpsi1}\\
\psi_{n,B}=g_{1}z_{1}^{n}\phi_{B}^{(1)}+g_{2}z_{2}^{n}\phi_{B}^{(2)}\label{gpsi2}
\end{align}
where $n=1,2,\cdots,N$. To solve the eigenvalue equation (\ref{BTeq}),
the general ansatz of wave function should also satisfy the boundary
conditions. Substituting Eqs.(\ref{gpsi1}) and (\ref{gpsi2}) into
Eq.(\ref{eq:Eboundry}), we obtain
\begin{equation}
M_{B}\left(\begin{array}{l}
g_{1}\\
g_{2}
\end{array}\right)=\left(\begin{array}{ll}
A\left(z_{1},N\right) & A\left(z_{2},N\right)\\
B\left(z_{1},N\right) & B\left(z_{2},N\right)
\end{array}\right)\left(\begin{array}{l}
g_{1}\\
g_{2}
\end{array}\right)=0
\end{equation}
with
\begin{align}
A(z_{i},N) & =\frac{P\mathrm{i}\gamma_{L}}{E_{P}}(J+hz_{i})-J,\label{ab}\\
B(z_{i},N) & =\frac{\mathrm{i}\gamma_{R}}{E_{P}}(h+Jz_{i})z_{i}^{N}-Jz_{i}^{N+1}.
\end{align}
The condition for the existence of nontrivial solutions of $\left(g_{1},g_{2}\right)$
is determined by $\operatorname{det}\left[M_{B}\right]=0$ \cite{GuoCX,Alase},
which leads to the following equation:
\begin{equation}
p_{1}\sin[N\theta]-p_{2}\sin[(N+1)\theta]+p_{3}\sin[(N-1)\theta]=0,\label{Atheta}
\end{equation}
where $p_{1}=\mathrm{i}J(P\gamma_{L}+\gamma_{R})E_{P}-(J^{3}-JP\gamma_{L}\gamma_{R})$,
$p_{2}=J^{2}h$, and $p_{3}=hP\gamma_{L}\gamma_{R}$.
By taking $\gamma_L=\gamma_R=\gamma$, we get Eq.(9) in the main text.
If $\theta$ is a solution of the above equation, it is clear that $-\theta$ should be also a solution of the equation. As both $\theta$ and $-\theta$ correspond to the same eigenvalue $E_P$, we can only consider one of them. Then, we can rewrite Eq.(\ref{Atheta})
in even channel ($P=1$) and odd channel ($P=-1$) as
\begin{equation}
\begin{array}{c}
[2\mathrm{i}\gamma E_{e}-(1-\gamma^{2})]\sin[N\theta]-h\sin[(N+1)\theta]\\
+h\gamma^{2}\sin[(N-1)\theta]=0
\end{array}\label{Aeq:thetaN1}
\end{equation}
and
\begin{equation}
(1+\gamma^{2})\sin[N\theta]+h\sin[(N+1)\theta]+ h\gamma^{2}\sin[(N-1)\theta]=0, \label{Aeq:thetaN2}
\end{equation}
respectively.

We note $T^P$ fulfills $K$ symmetry, i.e.,
\begin{equation}
\eta^{-1} T^P \eta = - (T^P)^*,
\end{equation}
where $\eta$ is a diagonal matrix with the elements $[-1,1,-1,...,(-1)^{2N}]$. The existence of $K$ symmetry implies that if $E$ is an
eigenvalue of $T^P$, then $-E^*$ is also an eigenvalue of $T^P$, i.e., both the even-parity and odd-parity rapidity spectra should distribute symmetrically about the imaginary axis.
For convenience, we shall use  $T^e$ and $T^{o}$ to denote  $T^P$ with $P=1$ (even parity) and $P=-1$ (odd parity), respectively.

For the even parity ($P=1$),  $T^e$ is invariant under the reflection operation, i.e.,
\begin{equation}
\mathscr{P} T^e \mathscr{P} = T^e.
\end{equation}
For the odd parity ($P=-1$),  $T^{o}$ fulfills the $\mathscr{P}\mathcal{T}$  symmetry, i.e.,
\begin{equation}
[\mathscr{P}\mathcal{T}, T^{o} ] = 0.
\end{equation}
The existence of $\mathscr{P}\mathcal{T}$ symmetry suggests that the eigenvalues of  $T^{o}$  are either real or distribute symmetrically about the real axis, i.e.,
if $E$ is an
eigenvalue of $T^{o}$, then $E^*$ is also an eigenvalue of $T^{o}$.


For convenience, we denote the eigenvalues as $E_{\pm}(\theta)=\pm\sqrt{J^{2}+h^{2}+2Jh\cos\theta}$, with $\Re[\sqrt{J^{2}+h^{2}+2Jh\cos\theta}]>0$.
While the solutions of $\theta$ are always complex for the even-parity case, they can be real or complex for the odd-parity case,  According to the expression of  $E_{\pm}(\theta)$, we have $E_{\pm}(\theta)^{*}=E_{\pm}(\theta^{*})$. If we assume that $E_{+}(\theta)$ is an eigenvalue of $T^{e}$, we can get $E_{-}(\theta^{*})$ is also the eigenvalue of $T^{e}$, i.e., when $\theta$ satisfies the Eq.(\ref{Aeq:thetaN1}) with $E_{e}=E_{+}$, then $\theta^{*}$ satisfies the Eq.(\ref{Aeq:thetaN1}) with $E_{e}=E_{-}$. According to Eq.(\ref{Aeq:thetaN1}), we can see that $-\theta^{*}$ is also the solution of Eq.(\ref{Aeq:thetaN1}) and  $E_{-}(-\theta^{*})= E_{-}(\theta^{*})$. In the odd-parity channel, the matrix $T^{o}$ has $\mathscr{P}\mathcal{T}$ symmetry and $K$ symmetry. We can get that when $E_{+}(\theta)$ is an eigenvalue of $T^{o}$, then $E_{+}(\theta^{*}), E_{-}(\theta), E_{-}(\theta^{*})$ are also the eigenvalues of $T^{o}$. So, when $\theta$ satisfies Eq.(\ref{Aeq:thetaN2}), then $\theta^{*}$ also satisfies Eq.(\ref{Aeq:thetaN2}).

\begin{figure*}[ht]
\center
\includegraphics[width=17cm]{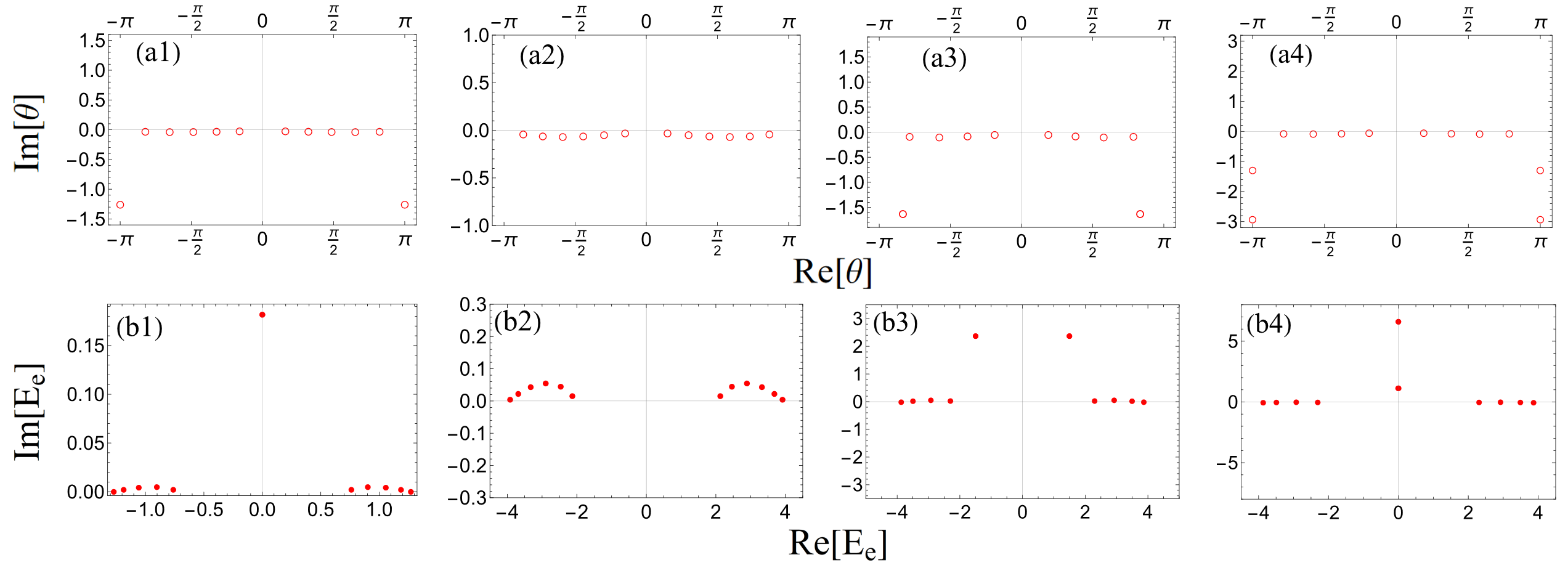} \caption{The solutions of Eq.(\ref{eq:thetaN1}) and the rapidity spectrum with $N=6$,
(a1), (b1) $J=1,h=0.3,\gamma=0.2$, (a2), (b2) $J=1,h=3,\gamma=0.2$
, (a3), (b3) $J=1,h=3,\gamma=5$ and (a4), (b4) $J=1,h=3,\gamma=8$.
The red empty circles represent the solutions of Eq.(\ref{eq:thetaN1}),
while the red points represent the rapidity spectrum from the even-parity channel.}
\label{appendixe}
\end{figure*}
\begin{figure*}[ht]
\center
\includegraphics[width=17cm]{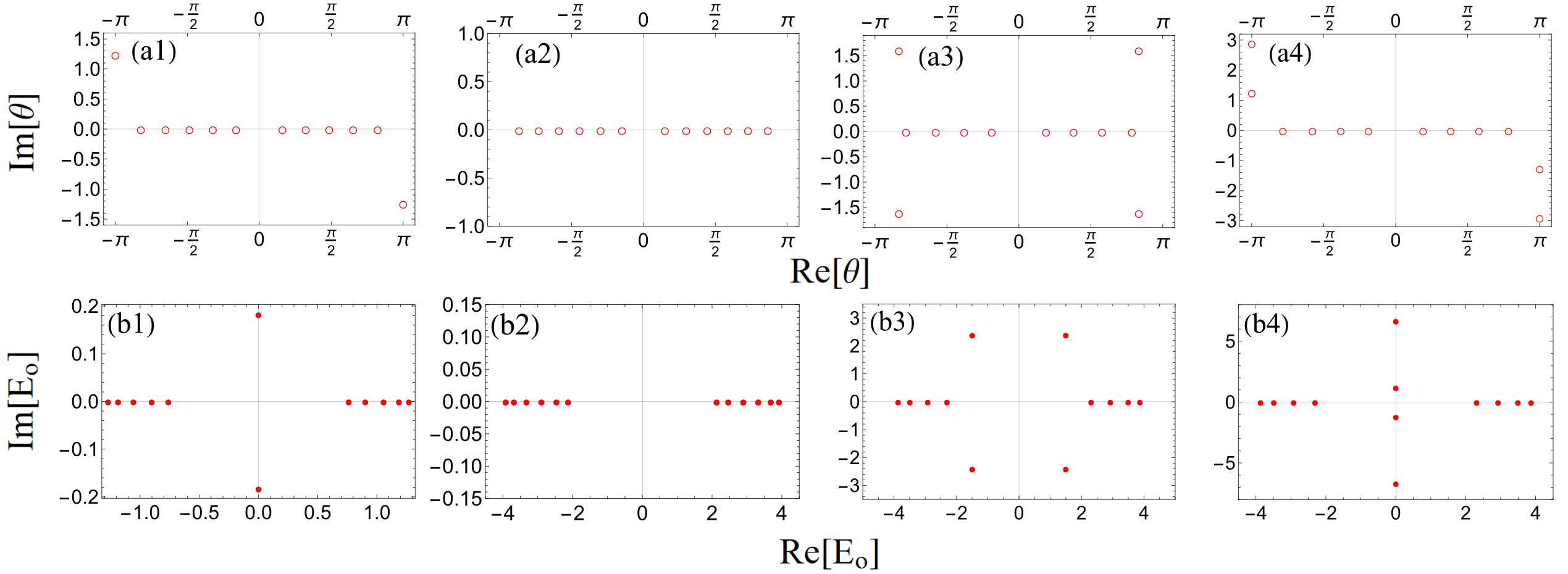} \caption{The solutions of Eq.(\ref{eq:thetaN2}) and the rapidity spectrum with $N=6$,
(a1), (b1) $J=1,h=0.3,\gamma=0.2$, (a2), (b2) $J=1,h=3,\gamma=0.2$
, (a3), (b3) $J=1,h=3,\gamma=5$ and (a4), (b4) $J=1,h=3,\gamma=8$.
The red empty circles represent the solutions of Eq.(\ref{eq:thetaN2}),
while the red points represent the rapidity spectrum from the odd-parity channel.}
\label{appendix}
\end{figure*}

To see clearly how the solutions of $\theta$ are related to the rapidity spectrum, in Fig.\ref{appendixe} and Fig.\ref{appendix} we demonstrate the solutions of Eq.(\ref{Aeq:thetaN1}) and Eq.(\ref{Aeq:thetaN2}), respectively and the corresponding even-parity and odd-parity rapidity spectrum for the four typical cases in Fig.\ref{FLS1}.
In Fig.\ref{appendixe}, the solutions satisfying with $\Re[\theta]>0$ correspond to $E_{+}$ and the solutions satisfying with $\Re[\theta]<0$ correspond to $E_{-}$. In Fig.\ref{appendixe} (a3), we note that the solutions of $\theta$ given by $2.61894-1.60997i$ and $2.61962-1.60891i$ are nearly degenerate. Similarly, the solutions given by  $-2.61894-1.60997i$ and $-2.61962-1.60891i$ are nearly degenerate. These solutions become exactly degenerate in the thermodynamical limit $N\rightarrow \infty$.

\section{Determine the boundary of the schematic phase diagram}
In the limit of $N \rightarrow \infty$, we can determine the boundary of schematic phase diagram by analyzing the solutions of Eq.(\ref{eq:thetaN2}) and the odd-parity rapidity spectrum.
We note that, in three-segment and nine-segment regions of the schematic phase diagram, the odd-parity rapidity spectrum has one pair and two pairs of pure imaginary eigenvalues, respectively. Next we look for conditions for the existence of pure imaginary rapidity eigenvalues. Without loss of generality, we assume the solution of Eq.(\ref{eq:thetaN2}) as $\theta=\theta_{R}+\mathrm{i}\theta_{I}$ and $\theta_{R}$, $\theta_{I}$ are purely real. Then, we can rewrite the corresponding eigenvalue as
\begin{equation}
E_{o}=\pm\sqrt{1+h^{2}+2h\cos[\theta_{R}+\mathrm{i}\theta_{I}]}.\label{AB1}
\end{equation}
If the eigenvalue $E_{o}$ is pure imaginary, we should have $\theta_{R}=\pi$, and thus the eigenvalue can be expressed as
\begin{equation}
E_{o}=\pm\sqrt{1+h^{2}-2h\cosh[\theta_{I}]}.\label{Ej-1}
\end{equation}

Substituting $\theta = \pi + \mathrm{i}\theta_{I}$ into Eq. (\ref{eq:thetaN2}), we can get
\begin{equation}\label{AB2}
  \begin{aligned}
& f_{1}\sinh\left[N\theta_{I}\right]-f_{2}\sinh\left[(N+1)\theta_{I}\right]
\\& -f_{3}\sinh\left[(N-1)\theta_{I}\right]=0,
  \end{aligned}
\end{equation}
where $f_{1}=\left(1+\gamma^{2}\right),f_{2}=h,f_{3}=h\gamma^{2}$.
In the thermodynamic limit of $N \rightarrow \infty$, we can use the formula:
\begin{equation}
2\sinh\left[N\theta_{I}\right]\thickapprox \textrm{sign}[\theta_{I}] e^{N |\theta_{I}|},\label{AB3}
\end{equation}
and thus the Eq.(\ref{AB2}) is simplified to
\begin{equation}
f_{1}e^{N|\theta_{I}|}-f_{2}e^{(N+1)|\theta_{I}|}-f_{3}e^{(N-1)|\theta_{I}|}=0,\label{AB4}
\end{equation}
which gives rise to
\begin{equation}
f_{1}e^{|\theta_{I}|}-f_{2}e^{2|\theta_{I}|}-f_{3}=0 .\label{AB5}
\end{equation}
Let $x=e^{|\theta_{I}|}$, and the solutions of $x$ are given by
 \begin{align}
x= \frac{1+\gamma^{2}\pm\sqrt{\left(1+\gamma^{2}\right)^{2}-4h^{2}\gamma^{2}}}{2h}. \label{AB6}
\end{align}
For convenience, we denote the above two solutions as $x_{\pm}=e^{|\theta_{I}(\pm)|}$.
For any nonzero solution of $\theta_{I}$, we always have $x>1$. If $x_{\pm}>1$, then we have two solutions $\theta_{I}(\pm)$, corresponding to two pairs of pure imaginary eigenvalues. If only one of $x_{\pm}$ is larger than $1$, then we have one pair of pure imaginary eigenvalues. If $x_{\pm}<1$, then no pure imaginary eigenvalue exits.

Next, we discuss in detail and set $h>0$, $\gamma>0$. When $x_{\pm}>1$, we get the constraint equations as follow,
\begin{equation}
\begin{aligned}
&\left(1+\gamma^{2}\right)^{2}-4h^{2}\gamma^{2}\geq0,\\
&\frac{1+\gamma^{2}+\sqrt{\left(1+\gamma^{2}\right)^{2}-4h^{2}\gamma^{2}}}{2h}>1,\\
&\frac{1+\gamma^{2}-\sqrt{\left(1+\gamma^{2}\right)^{2}-4h^{2}\gamma^{2}}}{2h}>1.
\end{aligned}
\end{equation}
Thus, we can get the boundary of nine-segment phase is $1<\gamma$ and $1<h\leq\frac{1+\gamma^{2}}{2\gamma}$.

For the three-segment phase, only one of $x_{\pm}$ is larger than $1$, and the constraint equations are as follow,
\begin{equation}
\begin{aligned}
&\left(1+\gamma^{2}\right)^{2}-4h^{2}\gamma^{2}\geq0,\\
&\frac{1+\gamma^{2}+\sqrt{\left(1+\gamma^{2}\right)^{2}-4h^{2}\gamma^{2}}}{2h}>1,\\
&\frac{1+\gamma^{2}-\sqrt{\left(1+\gamma^{2}\right)^{2}-4h^{2}\gamma^{2}}}{2h}\leq1.
\end{aligned}
\end{equation}
Here, it is easy to certify $x_{+}\geq x_{-}$. Then we can get the boundary of three-segment phase is $0<h<1$ (or $h=1$ and $\gamma>1$).

When $\gamma=1$, Eq.(\ref{eq:thetaN2}) reduces to
\begin{equation}
-2 \sin[N\theta]-h\sin[(N+1)\theta]-h\sin[(N-1)\theta]=0,
\end{equation}
which leads to
\begin{equation}
\sin[ N\theta](1+h \cos[\theta])=0.
\end{equation}
From the above equation, we see that the solutions are determined by
\begin{align}
  \sin[N\theta]=0,~~1+h \cos[\theta]=0,
\end{align}
which give rise to $\theta=\frac{j\pi}{N},
(j=1,\cdots,N)$ and $\theta=\arccos[-\frac{1}{h}]$. Here, $\theta=\pi$ should be abandoned, because $\theta=\pi$ corresponds to $z_1=z_2$ and thus Eq.(\ref{gpsi1}) and Eq.(\ref{gpsi2}) are not linearly independent.
When $h<1$, the equation $1+h \cos[\theta]=0$ has no purely real solution, and the solution is given by $\theta=\pi+\mathrm{i}\theta_{I}$ with $\theta_I=arcosh[1/h]$, corresponding to the existence of one pair of imaginary eigenvalues in the rapidity spectrum.
When $h>1$, $\theta=\arccos[-\frac{1}{h}]$ is purely real and the odd-parity rapidity spectrum is purely real. The line of $\gamma=1$ and $h>1$ is the phase boundary between the five-segment and one-segment regions. In the region of $h>1$ and $\gamma<1$, all solutions of Eq.(\ref{eq:thetaN2}) are real.


Next we analyze the boundary bound states of the even-channel rapidity spectrum in the thermodynamic limit of $N \rightarrow \infty$. In Fig.\ref{appendix3}(a2) and (b2), it is shown that the even-parity rapidity spectrum has one degenerate imaginary eigenvalue and two degenerate imaginary eigenvalues in three-segment and nine-segment regions of the schematic phase diagram, respectively. So, we assume the solution of Eq.(\ref{eq:thetaN1}) as $\theta=\theta_{R}+\mathrm{i}\theta_{I}$ and $\theta_{R}$, $\theta_{I}$ are purely real. Then, we can rewrite the corresponding eigenvalue as
\begin{equation}
E_{e}=\pm\sqrt{1+h^{2}+2h\cos[\theta_{R}+\mathrm{i}\theta_{I}]}.\label{ABc1}
\end{equation}
If the eigenvalue $E_{e}$ is pure imaginary, we should have $\theta_{R}=\pi$, and thus the eigenvalue can be expressed as
\begin{equation}
E_{e}=\pm\sqrt{1+h^{2}-2h\cosh[\theta_{I}]}.\label{Ej-2}
\end{equation}

Substituting $\theta = \pi + \mathrm{i}\theta_{I}$ into Eq. (\ref{eq:thetaN1}), we can get
\begin{equation}\label{ABc2}
\begin{aligned}
&4\gamma^2 E_{e}^2 \sinh[N \theta_{I}]^2+[(2-f_{1}) \sinh[N \theta_{I}]-
\\& f_{2}\sinh[(N+1)\theta_{I}]+f_{3}\sinh[(N-1)\theta_{I}]]^{2}=0 ,
\end{aligned}
\end{equation}
where $f_{1}=\left(1+\gamma^{2}\right),f_{2}=h,f_{3}=h\gamma^{2}$.
In the thermodynamic limit of $N \rightarrow \infty$, we can use the formula:
\begin{equation}
2\sinh\left[N\theta_{I}\right]\thickapprox \textrm{sign}[\theta_{I}] e^{N |\theta_{I}|},\label{AB3}
\end{equation}
and thus the Eq.(\ref{ABc2}) is simplified to
\begin{equation}
(f_{1}e^{N|\theta_{I}|}-f_{2}e^{(N+1)|\theta_{I}|}-f_{3}e^{(N-1)|\theta_{I}|})^2=0,\label{AB4}
\end{equation}
which gives rise to
\begin{equation}
f_{1}e^{|\theta_{I}|}-f_{2}e^{2|\theta_{I}|}-f_{3}=0 .\label{AB5}
\end{equation}
Let $x=e^{|\theta_{I}|}$, and the solutions of $x$ are given by
 \begin{align}
x= \frac{1+\gamma^{2}\pm\sqrt{\left(1+\gamma^{2}\right)^{2}-4h^{2}\gamma^{2}}}{2h}. \label{AB66}
\end{align}
It is easy to see that the results are the same as the analysis about Eq.(\ref{AB6}).

\section{Scaling relation of Liouvillian gap}

For the case of $\gamma_{L}=\gamma_{R}=\gamma$, the
Liouvillian gap comes from even channel, i.e., determined by the solutions of
\begin{equation}
\begin{aligned}
(\gamma^{2}-1+2\mathrm{i} & E_{e}\gamma)\sin[N\theta]+h\gamma^{2}\sin[(N-1)\theta]
\\&-h\sin[(N+1)\theta]=0. \label{appdeix-rs1}
\end{aligned}
\end{equation}
In general, solutions of the above equation are always complex. In the limits of weak and strong dissipation, we can derive analytical expression of Liouvillian gap by applying perturbation theory.

Firstly, we consider the weak dissipation limit with $\gamma\ll 1$ and $E_{e}=\sqrt{1+h^{2}+2h\cos[\theta]}$. By taking $\gamma$ as a small parameter for the perturbation calculation, the zero-order solution of  $\theta$ is determined by
\begin{equation}
\sin[N\theta^{(0)}]+h\sin[(N+1)\theta^{(0)}]=0.
\end{equation}
Thus we can get
\begin{equation}
\theta_{j}^{(0)}\simeq\left\{\begin{array}{l}
\frac{j\pi}{N}-\frac{h}{N}\sin[\frac{j\pi}{N}]~~(h\ll \frac{1}{N}),\\
\frac{2j\pi}{2N+1}~~(h=1),\\
\frac{j\pi}{N+1}+\frac{1}{h(N+1)}\sin[\frac{j\pi}{N+1}]~~(\frac{1}{h}\ll \frac{1}{N}),
\end{array}\right.
\end{equation}
where $j=1,2,\cdots,N$. Then we use the perturbation theory and assume $\theta_{j}=\theta_{j}^{(0)}+\gamma \theta_{j}^{(1)}$
with $\gamma\ll 1$. Substituting it into Eq.(\ref{appdeix-rs1}), we can get
\begin{equation}\label{D12}
\begin{aligned}
\theta_j^{(1)}&=-\frac{h \sin[ (N+1) \theta_{j}^{(0)}]+ \sin [N \theta_{j}^{(0)}]}{\left(h(N+1) \cos [(N+1) \theta_{j}^{(0)}]+ N \cos[ N \theta_{j}^{(0)}]\right) \gamma}\\&+\mathrm{i} \frac{2 E_{j,1}^{(0)} \sin [N \theta_{j}^{(0)}]}{h(N+1) \cos[(N+1) \theta_{j}^{(0)}]+ N \cos[ N \theta_{j}^{(0)}]},
\end{aligned}
\end{equation}
where $E_{e}^{(0)}=\sqrt{1+h^{2}+2h\cos[\theta_{j}^{(0)}]}$.
Thus, to the first order of $\gamma$, the spectrum can be approximately written as
\begin{equation}
\begin{aligned}
E_{e} & \approx  E_{e}^{(0)} \sqrt{1-\frac{2  h \gamma \theta_{j}^{(1)} \sin [\theta_j^{(0)}]}{(E_{e}^{(0)})^2}}\\& \approx E_{e}^{(0)}\left(1-\frac{ h \gamma \theta_{j}^{(1)} \sin \theta_{j}^{(0)}}{(E_{e}^{0})^2}\right)\\&=E_{e}^{(0)}-\frac{ h \theta_{j}^{(1)} \sin \theta_j^{(0)}}{E_{e}^{(0)}} \gamma,
\end{aligned}
\end{equation}
and the imaginary part of rapidity spectrum from the even channel can be written as
\begin{align}
\mathrm{I}_{j}\simeq |\frac{2\gamma h \sin[ \theta_j^{(0)}]\sin[ N \theta_j^{(0)}]}{ h(N+1) \cos[(N+1) \theta_j^{(0)}]+ N \cos [N \theta_j^{(0)}]}| .
\end{align}
We find that $\mathrm{I}_{j}$ with $j=1$ is the minimum, which is the same as the numerical result. So, we set $\theta^{(0)}=\theta_1^{(0)}$. Thus, the Liouvillian gap is given by
\begin{equation}
\begin{aligned}
\Delta_{g}&=4 \mathrm{I}_{1} \\&=|\frac{8\gamma h \sin[ \theta^{(0)}]\sin[ N \theta^{(0)}]}{h(N+1) \cos[(N+1) \theta^{(0)}]+N \cos [N \theta^{(0)}]}|.
\end{aligned}
\end{equation}

It follows that the scaling of the Liouvillian gap with lattice size $N$ is given by
\begin{equation}
\Delta_{g}\propto\gamma N^{-3},
\end{equation}
where $\sin[\theta^{(0)}]\propto N^{-1},\sin[N\theta^{(0)}]\propto N^{-1},\ \cos[N\theta^{(0)}]\propto N^{0}$ in the thermodynamic limit.

Now we consider the strong dissipation limit with $\gamma\gg 1$. We can rewrite Eq.(\ref{appdeix-rs1}) as
\begin{equation}\label{appdeix-rs9}
\begin{aligned}
(1-\frac{1}{\gamma^{2}}+&\frac{2\mathrm{i}E_{e}}{\gamma})\sin[N\theta]+h\sin[(N-1)\theta]
\\&-\frac{h}{\gamma^{2}}\sin[(N+1)\theta]=0.
\end{aligned}
\end{equation}
For $\gamma \gg 1$ and $E_{e}=\sqrt{1+h^{2}+2h\cos[\theta^{(0)}]}$, we can take $1/\gamma$ as a small parameter for perturbation calculation, and the zero-order solution of  $\theta_{j}$ is determined by
\begin{equation}
\sin[N\theta^{(0)}]+h\sin[(N-1)\theta^{(0)}]=0.
\end{equation}
Thus we can get
\begin{equation}
\theta_{j}^{(0)}\simeq\left\{\begin{array}{l}
\frac{j\pi}{N}+\frac{h}{N}\sin[\frac{j\pi}{N}]~~(h\ll \frac{1}{N}),\\
\frac{2j\pi}{2N-1}~~(h=1),\\
\frac{j\pi}{N-1}-\frac{1}{h(N-1)}\sin[\frac{j\pi}{N-1}]~~(\frac{1}{h}\ll \frac{1}{N}),
\end{array}\right.
\end{equation}
where $j=1,2,\cdots,N$. Then we use the perturbation theory and assume $\theta_{j}=\theta_{j}^{(0)}+\frac{1}{\gamma} \theta_{j}^{(1)}$
with $\gamma\gg 1$. Substituting it into Eq.(\ref{appdeix-rs9}), we can get
\begin{equation}\label{D11}
\begin{aligned}
\theta_j^{(1)}&=-\frac{(h \sin[ (N+1) \theta_{j}^{(0)}]+\sin [N \theta_{j}^{(0)}])\gamma}{h(N-1) \cos [(N-1) \theta_{j}^{(0)}]+ N \cos[ N \theta_{j}^{(0)}]}\\&-\mathrm{i} \frac{2 E_{j,1}^{(0)} \sin [N \theta_{j}^{(0)}]}{h(N-1) \cos [(N-1) \theta_{j}^{(0)}]+N \cos[ N \theta_{j}^{(0)}]}
\end{aligned}
\end{equation}
where $E_{e}^{(0)}=\sqrt{1+h^{2}+2h\cos[\theta_{j}^{(0)}]}$.

Thus, to the order of $1/\gamma$, the spectrum can be approximately represented as
\begin{equation}
\begin{aligned}
E_{e} &\approx E_{e}^{0} \sqrt{1-\frac{2 h \theta_j^{(1)} \sin [\theta_j^{(0)}]}{\gamma (E_{e}^{(0)})^2}} \\& \approx E_{e}^{0}\left(1-\frac{ h \gamma \theta_j^{(1)} \sin \theta_j^{(0)}}{\gamma (E_{e}^{0})^2}\right)
\\& =E_{e}^{(0)}-\frac{ h \theta_j^{(1)} \sin \theta_j^{(0)}}{E_{e}^{(0)} \gamma} ,
\end{aligned}
\end{equation}
and the imaginary part of rapidity spectrum from the even channel can be written as
\begin{equation}
\mathrm{I}_{j}\simeq |\frac{2h \sin[ \theta_j^{(0)}]\sin[ N \theta_j^{(0)}]}{(h(N-1) \cos [(N-1) \theta_{j}^{(0)}]+N \cos[ N \theta_{j}^{(0)}])\gamma }| .
\end{equation}
We find that $\mathrm{I}_{j}$ with $j=1$ is the minimum, which is the same as the numerical result. So, we set $\theta^{(0)}=\theta_1^{(0)}$. Thus, the Liouvillian gap is given by
\begin{equation}
\begin{aligned}
\Delta_{g}&=4 \mathrm{I}_{1}\\&=|\frac{8 h \sin[ \theta^{(0)}]\sin[ N \theta^{(0)}]}{(h(N-1) \cos [(N-1) \theta^{(0)}]+ N \cos[ N \theta^{(0)}])\gamma }|.
\end{aligned}
\end{equation}

It follows that the scaling of the Liouvillian gap with lattice size $N$  is given by
\begin{equation}
\Delta_{g}\propto\frac{ N^{-3}}{\gamma},
\end{equation}
where $\sin[\theta_{j}^{(0)}]\propto N^{-1},\sin[N\theta_{j}^{(0)}]\propto N^{-1},\ \cos[N\theta_{j}^{(0)}]\propto N^{0}$ in the thermodynamic limit.
We note that if we choose $E_{e}=-\sqrt{1+h^{2}+2h\cos[\theta^{(0)}]}$, we get the same result of Liouvillian gap.

\section{Duality relation of rapidity spectrum and Liouvillian gap in the thermodynamic limit}

The rapidity spectrum of the even-parity channel is determined by solving the following equation:
\begin{equation}\label{appdx-e}
\begin{aligned}
(\gamma^{2}-1+2\mathrm{i}\gamma &E_{e})\sin[N\theta]+h\gamma^{2}\sin[(N-1)\theta]
\\&-h\sin[(N+1)\theta]=0,
\end{aligned}
\end{equation}
where $E_{e}=\pm \sqrt{1+h^{2}+2h\cos[\theta]}$. Here, we should notice the solutions corresponding to two equations ($E_{\pm}(\theta_{\pm})=\pm \sqrt{1+h^{2}+2h\cos[\theta_{\pm}]}$) and denote the solutions of them as $\theta_{\pm}$, respectively, i.e.,
\begin{equation}\label{appdx-e1}
\begin{aligned}
(\gamma^{2}-1+2\mathrm{i}\gamma &E_{\pm}(\theta_{\pm}))\sin[N\theta_{\pm}]+h\gamma^{2}\sin[(N-1)\theta_{\pm}]
\\&-h\sin[(N+1)\theta_{\pm}]=0,
\end{aligned}
\end{equation}
where  $E_{\pm}(\theta_{\pm})=\pm \sqrt{1+h^{2}+2h\cos[\theta_{\pm}]}$. According to Eq.(\ref{appdx-e1}), if $\theta_{+}$ is a solution of the equation with eigenvalue $E_{+}(\theta_{+})$, then $\theta_{+}^{*}$ should be a solution of  the equation with eigenvalue $E_{-}(\theta_{-})$, i.e., we have $\theta_{-}(\lambda)=\theta_{+}^{*}(\lambda)$. Similarly, we can prove $\theta_{+}(\lambda)=\theta_{-}^{*}(\lambda)$. So we have $E_{-}(\theta_{-})=E_{-}(\theta_{+}^{*})=-E_{+}(\theta_{+}^{*})=-(E_{+}(\theta_{+}))^{*}$ and $E_{+}(\theta_{+})=-E_{-}(\theta_{-}^{*})=-(E_{-}(\theta_{-}))^{*}$, which are consistent with the requirement of the $K$ symmetry.

Firstly, we consider the case of $E_{+}$ and get the equation as follows,
\begin{equation}\label{appdx-e2}
\begin{aligned}
(\gamma^{2}-1+2\mathrm{i}\gamma & E_{+}(\theta_{+}))\sin[N\theta_{+}]+h\gamma^{2}\sin[(N-1)\theta_{+}]
\\&-h\sin[(N+1)\theta_{+}]=0.
\end{aligned}
\end{equation}
Then we consider the equation of $E_{-}$ with the dissipation strength $\gamma'$, and Eq.(\ref{appdx-e1}) can be rewritten as
\begin{equation}\label{appdx-e3}
\begin{aligned}
(\gamma'^{2}-1+2\mathrm{i}\gamma'& E_{-}(\theta_{-}))\sin[N\theta_{-}]+h\gamma'^{2}\sin[(N-1)\theta_{-}]
\\&-h\sin[(N+1)\theta_{-}]=0.
\end{aligned}
\end{equation}
By using the relation $E_{-}(\theta_{-})=-(E_{+}(\theta_{+}))^{*}=-E_{+}(\theta_{+}^{*})$ and $\theta_{-}=\theta_{+}^{*}$,  Eq.(\ref{appdx-e3}) can be rewritten as
\begin{equation}
\begin{aligned}
(\gamma'^{2}-1-2\mathrm{i}\gamma' & E_{+}(\theta_{+}^{*}))\sin[N\theta_{+}^{*}]+h\gamma'^{2}\sin[(N-1)\theta_{+}^{*}]
\\&-h\sin[(N+1)\theta_{+}^{*}]=0.
\end{aligned}
\end{equation}
Now we assume $\gamma'=\frac{1}{\gamma}$ and thus the above equation can be represented as follows,
\begin{equation}\label{appdx-e6}
\begin{aligned}
(\gamma^{2}-1+2\mathrm{i}\gamma & E_{+}(\theta_{+}^{*}))\sin[N\theta_{+}^{*}]+h\gamma^{2}\sin[(N+1)\theta_{+}^{*}]
\\&-h\sin[(N-1)\theta_{+}^{*}]=0.
\end{aligned}
\end{equation}

When the real part of $\theta_{+}$ is proportional to $\frac{1}{N}$ and the imaginary part of $\theta_{+}$ is proportional to $\frac{1}{N^{2}}$, we have $\sin[(N-1)\theta_{+}]\approx \sin[(N+1)\theta_{+}] $ in the thermodynamic limit. Comparing  Eq.(\ref{appdx-e2}) and Eq.(\ref{appdx-e6}), we can consider them to be the same in the thermodynamic limit.  Since Eq.(\ref{appdx-e3}) and Eq.(\ref{appdx-e6}) are equivalent, then we get $\theta_{+}(\gamma)\approx \theta_{-}^{*}(\frac{1}{\gamma})$ in the thermodynamic limit.
Using the relation $\theta_{-}^{*}(\frac{1}{\gamma})=\theta_{+}(\frac{1}{\gamma})$, we can get $\theta_{+}(\gamma)\approx\theta_{+}(\frac{1}{\gamma})$.  According to the above analysis, we have $E_{+}(\gamma)=E_{+}(\frac{1}{\gamma})$ in the thermodynamic limit. Similarly, we can get $E_{-}(\gamma)=E_{-}(\frac{1}{\gamma})$ in the thermodynamic limit. So, we can get
\begin{equation}
  E_{e}(\gamma)=E_{e}(\frac{1}{\gamma})
\end{equation}
in the thermodynamic limit. Notice that for the boundary bound states, the real parts of them are not proportional to $\frac{\pi}{N}$ and thus the approximation $\sin[(N-1)\theta_{+}]\approx \sin[(N+1)\theta_{+}]$ does not hold true.

According to the definition of Liouvillian gap,
it follows that the Liouvillian gap fulfills
\begin{equation}
  \Delta_{g}(\gamma,h)= \Delta_{g}(\frac{1}{\gamma},h).
\end{equation}

\begin{figure}[ht]
\centering\includegraphics[width=8.5cm]{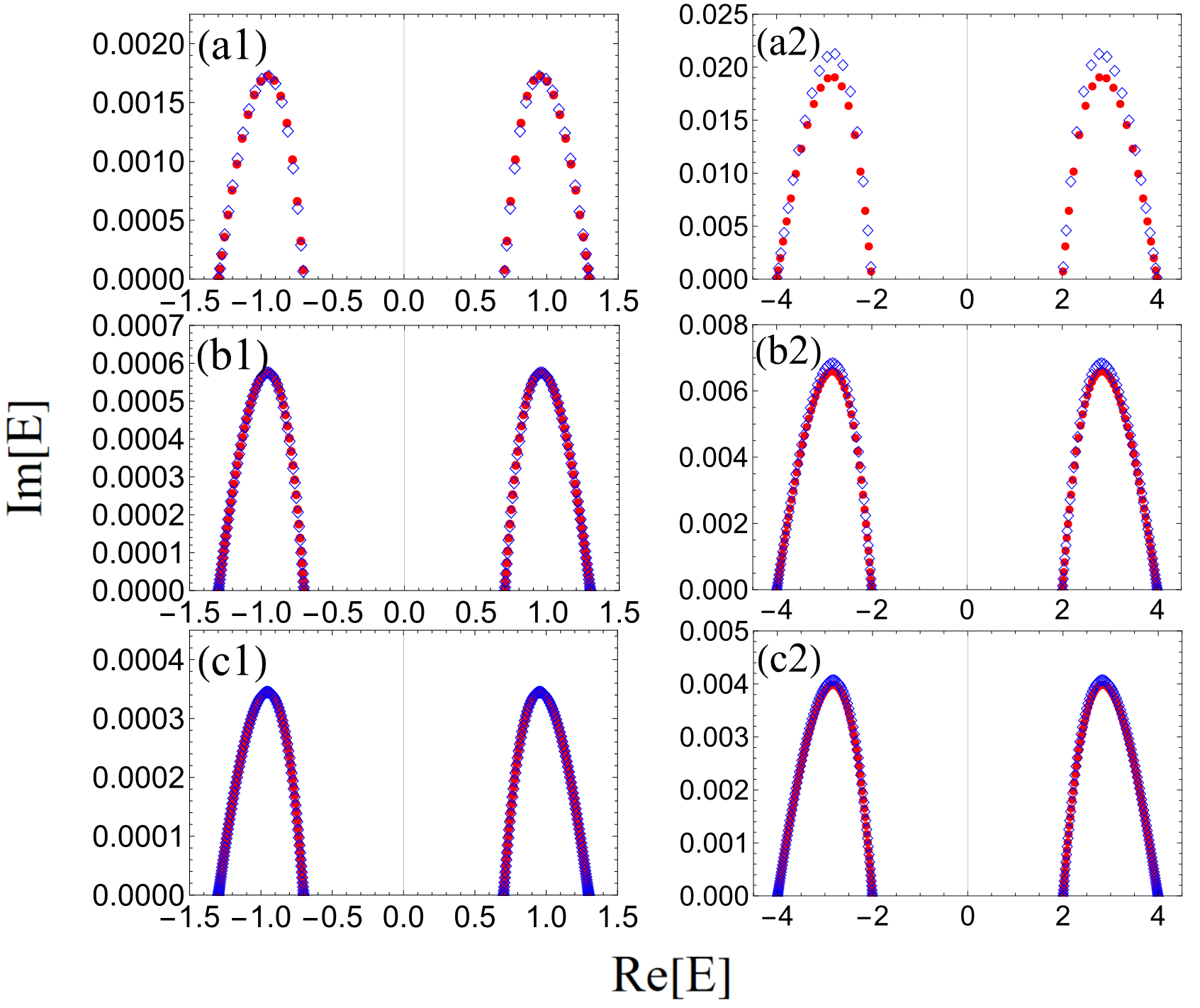} \caption{The rapidity spectrum from even channels (except for  eigenvalues corresponding to the boundary bound states) with $N=20$, (a1) $h=0.3$,
and (a2) $h=3$ , and $N=60$, (b1) $h=0.3$,
and (b2) $h=3$, and $N=100$, (c1) $h=0.3$,
and (c2) $h=3$. We take parameter $\gamma=0.2$ which corresponds to the red points and $\gamma=5$ which corresponds to the blue empty prismatic.}
\label{evenchannel}
\end{figure}

We have proven that the most of rapidity spectrum (except for those corresponding to the boundary bound states) satisfies the duality relation  $E(\gamma)=E(\frac{1}{\gamma})$ in the thermodynamic limit. The eigenvalues of boundary bound states do not satisfy the duality relation, e.g., the imaginary part of the boundary bound state (the one with larger imaginary part) increases with the increase of $\gamma$ as displayed in Fig.\ref{appendix3} (b3). In Fig.\ref{evenchannel}, we show the rapidity spectrum of the even channel except for the boundary bound state.  While the the number of bound states in the weak and strong dissipation regions are the same in the region of $h<1$, they are different in the region of $h>1$. This leads to the spectrum $E(\gamma)$ and $E(1/\gamma)$ in the region of $h=0.3$ coinciding much better than in the region of $h=3$, as shown in Fig.\ref{evenchannel}(a1) and (a2) for $N=20$. With the increase of lattice size,  the spectrum $E(\gamma)$ and $E(1/\gamma)$ coincide much better. While the spectrum $E(\gamma)$ and $E(1/\gamma)$ do not coincide very well for the case of $h=3$ with $N=20$, they are almost the same with $N=100$.

\section{Calculation of the average magnetization by using Lyapunov equation method}
Here we show the details of calculation of the average magnetization by using Lyapunov equation method \cite{zhang2022arxiv}.  Lyapunov equation method enables us to calculate the dynamics of systems with large sizes. In terms of the representation of Majorana fermion,  we can rewrite the transverse field Ising chain and the dissipation operators as $\hat{H}=\sum_{i,j}\hat{w}_{i}H_{i,j}\hat{w}_{j},\ \hat{L}_{\mu}=\sum_{j}l_{\mu,j}\hat{w}_{j}$. The matrix $\Gamma$ is also defined as Majorana fermion $\Gamma_{j,k}=\mathrm{i}\left\langle \hat{w}_{j}\hat{w}_{k}\right\rangle -\frac{\mathrm{i}}{2}\delta_{j,k}$. And then, we have
\begin{equation}
\partial_{t}\Gamma=X\Gamma+\Gamma X^{T}+Y,\label{eq:Gamma}
\end{equation}
where the matrix $X$, $Y$ are defined as $X=-2\mathrm{i}H-\Re(\sum_{\mu}l_{\mu}l_{\mu}^{\dagger}),\ Y=\Im(\sum_{\mu}l_{\mu}l_{\mu}^{\dagger})$, respectively. And $l_{\mu}$ can be expressed as $l_{\mu}=[l_{\mu,1},l_{\mu,2},\cdots l_{\mu,2N}]^{T}$.

Firstly, we apply the Jordan-Wigner transformation
\begin{equation}\label{jwappendix}
 \hat{a}_{j}^{\dagger}=\frac{1}{2}(\sigma_{j}^{x}+\mathrm{i}\sigma_{j}^{y})\prod_{l=1}^{j-1}
 (\sigma_{l}^{z}). ~~(1\leq j \leq N)
\end{equation}
Here, $\hat{a}_j$ and $\hat{a}_j^{\dagger}$ obey the canonical anti-commutation relations
$
 \left\{\hat{a}_i, \hat{a}_j^{\dagger}\right\} \equiv \hat{a}_i \hat{a}_j^{\dagger}+\hat{a}_j^{\dagger} \hat{a}_i=\delta_{i, j} \quad \text { and } \quad\left\{\hat{a}_i, \hat{a}_j\right\}= \left\{\hat{a}_i^{\dagger}, \hat{a}_j^{\dagger}\right\}=0.
$
And then, the transverse field Ising chain can be written as fermion form
\begin{equation}\label{ishappendix}
H=J\sum_{j=1}^{N-1}(\hat{a}_{j}^{\dagger}- \hat{a}_{j})(\hat{a}_{j+1}^{\dagger}+\hat{a}_{j+1})-h\sum_{j=1}^{N}(\hat{a}_{j}^{\dagger}- \hat{a}_{j})(\hat{a}_{j}^{\dagger}+\hat{a}_{j}).
\end{equation}
Then, we employ the self-adjoint Majorana operators $\hat{w}_{j,\pm}=\left(\hat{w}_{j,\pm}\right)^{\dagger}$\cite{zhang2022arxiv}
\begin{align}
\left(\begin{array}{c}\hat{w}_{j,+} \\ \hat{w}_{j,-}\end{array}\right):=\frac{1}{\sqrt{2}}\left(\begin{array}{cc}1 & 1 \\ \mathrm{i} & -\mathrm{i}\end{array}\right)\left(\begin{array}{c}\hat{a}_j \\ \hat{a}_j^{\dagger}\end{array}\right).
\end{align}
And we get
\begin{equation}
\hat{H}=2\mathrm{i}(J\sum_{j=1}^{N-1}\hat{w}_{j,-}\hat{w}_{j+1,+}
+h\sum_{j=1}^{N}\hat{w}_{j,+}\hat{w}_{j,-})
\end{equation}

For convenience, we denote
\begin{equation}
\hat{w}_j:=\hat{w}_{j,+} \text { and } \hat{w}_{j+N}:=\hat{w}_{j,-}.
\end{equation}
The Majorana operators with $j=1, \ldots, N$. They obey the anti-commutation relations
\begin{equation}
\left\{\hat{w}_i, \hat{w}_j\right\}=\delta_{i, j} \quad \text { for } \quad i, j=1, \ldots, 2 N.
\end{equation}
Thus, we get the transverse field Ising chain as
\begin{equation}
\hat{H}=\sum_{i, j} \hat{w}_i H_{i, j} \hat{w}_j.
\end{equation}
And the boundary dissipation operators can be rewritten as
\begin{equation}
  \hat{L}_{L}=\sqrt{2\gamma} \hat{w}_{1,+} , \hat{L}_{R}= \mathrm{i}Q\sqrt{2\gamma}\hat{w}_{N,-} .
\end{equation}
where $Q=\prod_{j=1}^{N}\sigma_{j}^{z}$ is the parity operator. In our case, the initial state is the even fermion state and the operator $m^{z}$ is also even fermion. Thus the distribution of the odd channel is zero, so we can only consider the average magnetization in the even channel i,e, $Q=1$. Thus, the matrix $X,\ Y$ is represented in terms of a $2N\times2N$ non-hermitian matrix as follow
\begin{equation}
\begin{array}{c}
X=\left[\begin{array}{cccc|cccc}
-2\gamma &  &  &  & 2h\\
 & 0 &  &  & -2J & \ddots\\
 &  & \ddots &  &  & \ddots & \ddots\\
 &  &  & 0 &  &  & -2J & 2h\\
\hline -2h & 2J &  &  & 0\\
 & \ddots & \ddots &  &  & \ddots\\
 &  & \ddots & 2J &  &  & 0\\
 &  &  & -2h &  &  &  & -2\gamma
\end{array}\right], \\
\\
  Y=0_{2N}.
\end{array}
\end{equation}
And then, we give the spectral representation of $X$
\begin{equation}
X=\sum_{j=1}^{2N}s_{j}\left|\psi_{jR}\right\rangle \left\langle \psi_{jL}\right|,
\end{equation}
and get Eq.{[}\ref{eq:Gamma}{]} as
\begin{equation}
\Gamma(t)=\sum_{j,k=1}^{2N}\left[e^{(s_{j}+s_{k})t}\left\langle \psi_{jL}\right|\Gamma(0)\left|\psi_{kL}\right\rangle \left|\psi_{jR}\right\rangle \left\langle \psi_{kR}\right|\right],
\end{equation}
in which $\left|\psi_{kL}\right\rangle =\left[\left\langle \psi_{kL}\right|\right]^{T},\ \left\langle \psi_{kR}\right|=\left[\left|\psi_{kR}\right\rangle \right]^{T}$.
So, the average magnetization is
\begin{equation}
\begin{aligned}
\langle m^{z}(t)\rangle & =\langle\frac{1}{N}\sum_{j=1}^{N}\sigma_{j}^{z}(t)\rangle=-\frac{2\mathrm{i}}{N}\sum_{j=1}^{N}\langle\hat{w}_{j,+}\hat{w}_{j,-}\rangle\\
&=\frac{1}{N}\sum_{j=1}^{N}(\Gamma_{N+j,j}-\Gamma_{j,N+j})\\
 & =\frac{1}{N} \text{Tr} \left(\left[\begin{array}{cc}
0_{N} & I_{N}\\
-I_{N} & 0_{N}
\end{array}\right]\Gamma(t)\right),
\end{aligned}
\end{equation}
where the matrix $\Gamma$ of the initial state of all spin up is denoted as
\begin{equation}
\Gamma(0)=\left[\begin{array}{cc}
0_{N} & -I_{N}/2\\
I_{N}/2 & -\mathrm{i}I_{N}
\end{array}\right].
\end{equation}

\begin{figure*}[htb]
\flushleft \includegraphics[width=17cm]{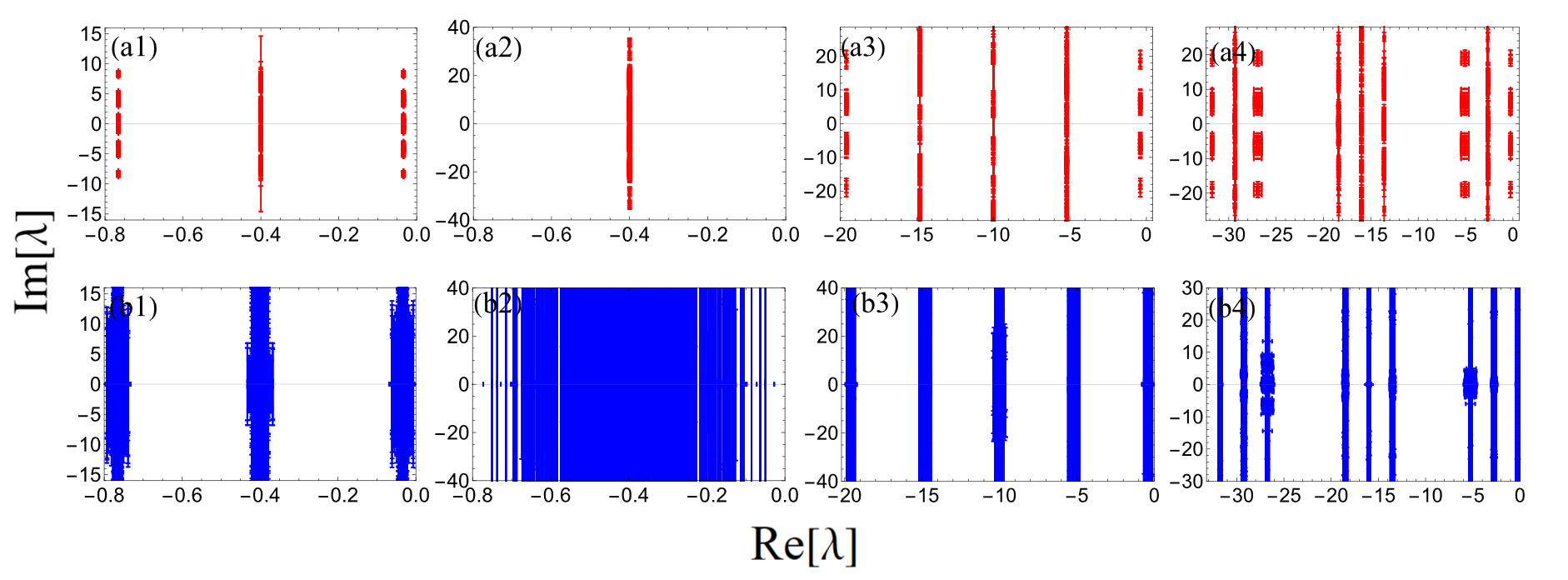} \caption{ The Liouvillian spectrum of odd channel (a1-a4) and even channel (b1-b4) with disorder strength $\delta=0.1$ for systems with $N=6$, and
(a1), (b1) $h=0.3,\gamma=0.2$, and (a2), (b2) $h=3,\gamma=0.2$
, and (a3), (b3) $h=3,\gamma=5$ and (a4), (b4) $h=3,\gamma=8$. The data are obtained by taking over $50$ random configurations.}
\label{appendixdis}
\end{figure*}

\section{Stability of stripe structure against random on-site disorder perturbation}

Here we demonstrate that the stripe structures are stable against random on-site disorder perturbation in the transverse field. The boundary-dissipated transverse field Ising model with transverse field strength being perturbed by random on-site disorder is described by
\begin{align}
H=-J\sum_{j=1}^{N-1}\sigma_{j}^{x}\sigma_{j+1}^{x}-h_{j}\sum_{j=1}^{N}\sigma_{j}^{z},\label{Hd}
\end{align}
where $h_{j}=h(1+\delta_j)$ $(j=1,2,\cdots,N)$ is uniformly distributed in the interval $[h(1-\delta),h(1+\delta)]$, i.e., $\delta_j$ is a random number uniformly distributed in $(-\delta ,\delta)$.

To see clearly how the disorder changes the structure of Liouvillian spectrum, we display the Liouvillian spectrum from the odd and even channels in Fig.\ref{appendixdis} (a1-a4) and (b1-b4), respectively. To compare with Fig.\ref{FLS1}, we set the same parameters and the strength of the disorder as $10\%$ ($\delta=0.1$). It is shown that the structures of the stripes are still discernable even we introduce random on-site disorder, i.e, the spectrum structure is stable against random on-site disorder perturbation.

\end{document}